\documentclass[12pt,twoside]{article}

\usepackage{epsf}
\usepackage{cite}

\usepackage{times}
\usepackage{fancyheadings}
\advance \headheight by 3.0truept       % for 12pt mandatory...

\renewcommand{\thefootnote}{\fnsymbol{footnote}}

\def\babar{\mbox{\sl B\hspace{-0.4em} {\scriptsize\sl
A}\hspace{-0.4em} B\hspace{-0.4em} {\scriptsize\sl A\hspace{-0.1em}R}}}

\newcommand{\lt}{\left}
\newcommand{\rt}{\right}
\newcommand{\ov}{\overline}
\newcommand{\imag}{{\rm Im}\,}
\renewcommand{\Im}{\imag}
\newcommand{\real}{{\rm Re}\,}
\renewcommand{\Re}{\real}

\newcommand{\nn}{\nonumber \\}
\newcommand{\no}{\nonumber }
\def\openone{\leavevmode\hbox{\small1\kern-3.8pt\normalsize1}}%

\newcommand{\eq}[1]{(\ref{#1})}

\newcommand{\ds}{\displaystyle}
\newcommand{\bra}[1]{\langle \, #1 \, | }
\newcommand{\ket}[1]{| \, #1 \, \rangle }

\newcommand{\bbd}{$B_d$--$\ov{B}{}_d\,$}
\newcommand{\bbs}{$B_s$--$\ov{B}{}_s\,$}
\newcommand{\bbms}{$B_s$--$\ov{B}{}_s\,$\ mixing}
\newcommand{\bbmd}{$B_d$--$\ov{B}{}_d\,$\ mixing}

\newcommand{\dbo}{\ensuremath{|\Delta B|  = 1}}

\newcommand{\dm}{\ensuremath{\Delta m}}
\newcommand{\dg}{\ensuremath{\Delta \Gamma}}

\newcommand{\adi}{{\cal A}_{\rm CP}^{\rm dir}}
\newcommand{\ami}{{\cal A}_{\rm CP}^{\rm mix}}
\newcommand{\adg}{{\cal A}_{\rm \Delta\Gamma}}

\newcommand{\gtf}{\ensuremath{\Gamma (B_s (t) \rightarrow f )}}
\newcommand{\gbtf}{\ensuremath{\Gamma (\ov{B}_s (t) \rightarrow f )}}
\newcommand{\gtfb}{\ensuremath{\Gamma (B_s(t) \rightarrow \ov{f} )}}
\newcommand{\gbtfb}{\ensuremath{\Gamma (\ov{B}_s(t) \rightarrow \ov{f} )}}
\newcommand{\Bsun}{\ensuremath{B_s^\textrm{\scriptsize  un}}}

\newcommand{\gunt}[1]{\ensuremath{ \Gamma \lt[#1,t\rt] }}
\newcommand{\guntf}{\ensuremath{ \Gamma  [f,t] }}
\newcommand{\guntfb}{\ensuremath{\Gamma  [\ov{f},t] }}
\newcommand{\guntfcpp}{\ensuremath{\Gamma [ f_{\rm CP+} ,t] }}
\newcommand{\brunt}[1]{ \ensuremath{Br [ { #1} ]}}

\newcommand{\epm}[2]{
 \raisebox{-0.5ex}{\shortstack[l]{$\scriptstyle+#1$\\$\scriptstyle-#2$}}}

\newcommand{\prd}{Phys.\ Rev.~D}
\newcommand{\plb}{Phys.\ Lett.~B}

\newlength{\nseparation}
\newenvironment{nfigure}
        {\begin{figure}[tb]\hrule\vspace{\nseparation}\par}
        {\vspace{\nseparation}\par \hrule \end{figure}}
\newenvironment{ntable}
        {\begin{table}[tb]\hrule\vspace{\nseparation}\par}
        {\vspace{\nseparation}\par \hrule \end{table}}

\begin{document}

%%%%%%%%%%% Titlepage

\begin{titlepage}
\parbox[t]{6cm}{
Fermilab--Pub--00/245-T \\ 
{ DESY 00--171}\\
CERN-TH/2000-333}
\hfill
\parbox[t]{4cm}{December 2000\\
hep-ph/0012219}\\
\vskip1.5truecm
\begin{center}
\boldmath
{\Large \bf In Pursuit of New Physics with $B_s$ Decays}
\unboldmath

\vspace{1cm}
{\sc Isard~Dunietz}${}^{1,}$\footnote{E-mail: dunietz@fnal.gov},
{\sc Robert Fleischer}${}^{2,}$\footnote{E-mail: Robert.Fleischer@desy.de}
and
{\sc Ulrich Nierste}${}^{3,}$\footnote{E-mail: Ulrich.Nierste@cern.ch}
\\[0.5cm]
\vspace*{0.1cm} ${}^1${\it Fermi National Accelerator Laboratory, 
                           Batavia, IL 60510-500, USA
}\\[0.3cm]
\vspace*{0.1cm} ${}^2${\it Deutsches Elektronen-Synchrotron DESY, 
Notkestr.\ 85, D--22607 Hamburg, Germany
}\\[0.3cm]
\vspace*{0.1cm} ${}^3${\it  CERN Theory Division, 1211 Geneva 23,
Switzerland
}\\[0.8cm]

%\vspace*{0.6cm}

%{\em Version of \today}

\vspace*{1truecm}

{\large\bf Abstract\\[10pt]} 
\parbox[t]{\textwidth}{ 
  The presence of a sizeable CP-violating phase in \bbms\ { would be}
  an unambiguous signal of physics beyond the Standard Model.  We
  analyse various possibilities to detect such a new phase considering
  both tagged and untagged decays. The effects of a sizeable
    width difference \dg\ between the $B_s$ mass eigenstates, on which
    the untagged analyses rely, are included in all formulae.  A
  novel method to find this phase from simple measurements of
  lifetimes and branching ratios in untagged decays is proposed. This
  method does not involve two-exponential fits, which require much
  larger statistics.  For the tagged decays, an outstanding role is
  played by the observables of the time-dependent angular distribution
  of the $B_s\to J/\psi[\to l^+l^-]\, \phi[\to K^+K^-]$ decay
  products. We list the formulae needed for the angular analysis in
  the presence of both a new CP-violating phase and a sizeable \dg, 
  and propose methods to remove a remaining discrete
  ambiguity in the new phase. This phase can therefore be
  determined in an unambiguous way.
}
\end{center}

\end{titlepage}

\thispagestyle{empty}
\vbox{}
\newpage
 
\setcounter{page}{1}

\setcounter{footnote}{0}
\renewcommand{\thefootnote}{\arabic{footnote}}

\section{Introduction}\label{sec:intro}
The rich phenomenology of non-leptonic $B$ decays offers various
strategies to explore the phase structure of the
Cabibbo--Kobayashi--Maskawa (CKM) matrix \cite{revs} and to search for
manifestations of physics beyond the Standard Model
\cite{new-phys}. Concerning the latter aspect, CP violation in \bbms\
is a prime candidate for the discovery of non-standard physics. In the
first place the \bbms\ amplitude is a highly CKM-suppressed
loop-induced fourth order weak interaction process and therefore very
sensitive to new physics.  Moreover in the Standard Model the
mixing-induced CP asymmetries in the dominant $B_s$ decay modes
practically vanish, because they are governed by the tiny phase $\arg
(-V_{tb} V_{ts}^*/(V_{cb} V_{cs}^*))$. It does not take much new
physics to change this prediction: already a fourth fermion
generation\footnote{ This scenario is still possible, though somewhat
disfavoured by electroweak precision data \cite{s}.}  can easily lead
to a sizeable new CP-violating phase in \bbms\ \cite{gnr}. It is
further possible that there are new flavour-changing interactions
which do not stem from the Higgs-Yukawa sector. The phases of these
couplings are not related to the phases of the CKM elements and
therefore induce extra CP violation. An example is provided by generic
supersymmetric models in which new flavour-changing couplings come
from off-diagonal elements of the squark mass matrix
\cite{ggms}. While such new contributions are { likely to affect also}
\bbmd, they appear in the $B_d$ system as a correction to a non-zero
Standard Model prediction for the mixing-induced CP asymmetry, which
involves the poorly known phase $\beta=\arg (-V_{tb} V_{td}^*/(V_{cb}
V_{cd}^*))$. To extract the new physics here additional information on
the unitarity triangle must be used.  In the $B_s$ system, however,
the new physics contribution is a correction to essentially zero
\cite{n}.

Indeed, the discovery of new physics through a non-standard  
CP-violating phase in \bbms\ may be achievable before the LHCb/BTeV era,
in Run-II of the Fermilab Tevatron. 

$B_s$-meson decays into final CP eigenstates that are caused by $\bar
b\to \bar c c \bar s$ quark-level transitions such as $B_s\to
D_s^+D^-_s$, $J/\psi\, \eta^{(\prime)}$ or $J/\psi\, \phi$, are
especially interesting \cite{nirsil,silver,bfNP}.  The $\eta$ and
$\eta^\prime$ mesons in $B_s\to J/\psi\, \eta^{(\prime)}$ can be
detected through $\eta\to\gamma\gamma$ and $\eta^\prime\to\rho^0
\gamma$, $\pi^+ \pi^- \eta$, or through $\eta\to\pi^+ \pi^- \pi^0$
\cite{hc}.  These modes require photon detection.  In the case of
$B_s\to J/\psi[\to l^+l^-]\, \phi[\to K^+K^-]$, which is particularly
interesting for $B$-physics experiments at hadron machines because of
its nice experimental signature, the final state is an admixture of
different CP eigenstates. In order to disentangle them, an angular
analysis has to be performed \cite{ddlr,ddf1}.  Experimental attention
is also devoted to three-body final states \cite{pm}. $B_s$-meson
decays triggered by the quark decay $\bar b \to \bar c u\bar d$ can
likewise access a CP-specific final state, e.g.\ via $B_s \to
D^0_{\rm CP+}[\to K^+ K^-] K_S$, with a likewise negligibly small
CP-violating phase in the Standard Model. The key point here is that
there are many different decay modes which all contain the same
information on the pursued new CP-violating phase $\phi$. Furthermore,
additional information on $\phi$ can be gained from analyses 
that require no tagging.  Untagged studies determine
$|\cos \phi|$ and are superior to tagged analyses in terms of
efficiency, acceptance and purity. However, they require a sizeable
width difference $|\dg|$ between the $B_s$ mass eigenstates. On the
other hand, from tagged analyses {(such as CP asymmetries)} $\sin \phi$
can be extracted, if the rapid \bbs\ oscillation can be resolved.
Both avenues should be pursued and their results combined, because
they measure the same fundamental quantities.

If we denote the Standard Model and the new physics contributions to
the \bbms\ amplitude with $S_{\rm SM}$ and $S_{\rm NP}$, respectively,
then the measurement of the mass difference \dm\ in the $B_s$ system
determines $|S_{\rm SM}+S_{\rm NP}|$. The knowledge of both \dm\ and
the \bbms\ phase $\phi$ then allows to solve for both the magnitude
and phase of $S_{\rm NP}$. Information on $\phi$ is especially
valuable, if $|S_{\rm SM}|$ and $|S_{\rm NP}|$ are comparable in size
and $\dm$ agrees within a factor of 2 or 3 with the Standard Model
prediction.

The purpose of this paper is twofold: we first identify useful
measurements and show how the information from different decay modes
and different observables can be combined in pursuit of a
statistically significant ``smoking gun'' of new physics. Second we
show how the \bbms\ phase can be identified unambiguously, without
discrete ambiguities.  The outline is as follows: after setting up our
notation in Section~\ref{sec:p} we consider untagged $B_s$ decays and
discuss various methods to determine $|\cos \phi|$ in
Section~\ref{sec:untagged}.  { Tagged $B_s$ decays are} discussed in
Section~\ref{sec:tagged}, whereas Section~\ref{sec:ambig} shows how to
resolve the discrete ambiguity in $\phi$.  Finally,
we conclude in Section~\ref{sec:concl}.
\section{Preliminaries}\label{sec:p}
In this section we define the various quantities entering the time
evolution of $B_s$ mesons and their decay amplitudes. We closely
follow the notation of the \babar-Book \cite{revs}.  Some of the
discussed quantities depend on phase conventions and enter physical
observables in phase-independent combinations \cite{d}. Since
this feature is well understood and extensively discussed in the
standard review articles \cite{revs}, we here fix some of these phases 
for convenience and only briefly touch this issue where necessary.

We choose the following convention for the CP transformation of meson
states and quark currents:\footnote{metric $g_{\mu \nu}=(1,-1,-1,-1)$}
\begin{eqnarray}
CP \ket{B_s} = - \ket{\ov{B}{}_s}, \qquad 
CP \, \ov{q}_L \gamma_\mu b_L \, (CP)^{-1} \;=\; 
        { - \ov{b}_L \gamma^\mu q_L } 
        . \label{defcp}
\end{eqnarray}
Hence the CP eigenstates are 
\begin{eqnarray}
\ket{B_s^{\textrm{\scriptsize even}}} &=& 
 \frac{1}{\sqrt{2}} \lt( \ket{B_s} - \ket{\ov{B}{}_s} \rt), \qquad 
 \mbox{and} \qquad
\ket{B_s^{\textrm{\scriptsize odd}}} \;=\; 
 \frac{1}{\sqrt{2}} \lt( \ket{B_s} + \ket{\ov{B}{}_s} \rt) 
 .\label{cpe}  
\end{eqnarray}
The time evolution of the \bbs\ system is governed by a Schr\"odinger
equation:
\begin{eqnarray}
i\, \frac{d}{d t} \pmatrix{ \ds \ket{B_s(t)} \cr
  \ds \ket{\ov{B}_s(t)} \cr } 
&=& \left( M - i\, \frac{\Gamma}{2} \right) \pmatrix{ \ds \ket{B_s(t)} \cr
  \ds \ket{\ov{B}_s(t)} \cr } \label{schr}
\end{eqnarray} 
with the mass matrix { $M=M^\dagger$} and the decay matrix 
{ $\Gamma=\Gamma^\dagger$}. Here
$\ket{B_s(t)}$ denotes the state of a meson produced as a $B_s$ at time
$t=0$, with an analogous definition for $\ket{\ov{B}_s(t)}$.  The
off-diagonal elements $M_{12}=M_{21}^*$ and
$\Gamma_{12}=\Gamma_{21}^*$ correspond to \bbms.  In the Standard
Model the leading contributions to $M_{12}$ and $\Gamma_{12}$ stem
from the box diagram in Fig.~\ref{fig:box};
\begin{nfigure}
\centerline{\epsfysize=4cm \epsffile{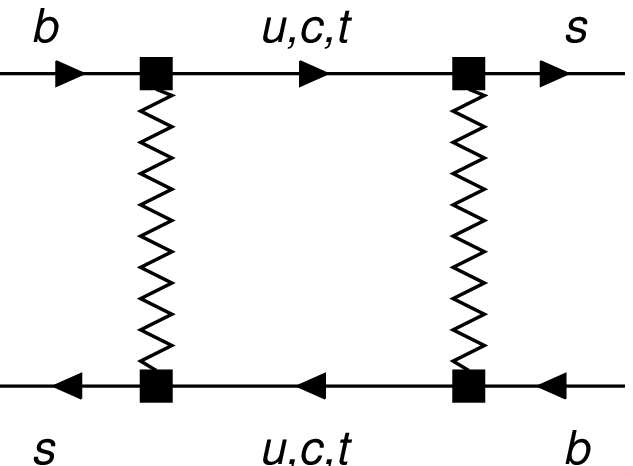}}
\caption{  \bbms\ in the Standard Model.}\label{fig:box}
\end{nfigure}
$\Gamma_{12}$ originates from the real final states into which both
$B_s$ and $\ov{B}_s$ can decay. It receives contributions from
box diagrams with light $u$ and $c$ quarks. Since $\Gamma_{12}$ is
dominated by { CKM-favoured} tree-level decays, it is practically
insensitive to new physics. On the other hand, $M_{12}$ is almost completely
induced by short-distance physics. Within the Standard Model the top
quarks in Fig.~\ref{fig:box} give the dominant contribution to
\bbms. This contribution is suppressed by four powers of the weak
coupling constant and two powers of $|V_{ts}|\simeq 0.04$.  Hence new
physics can easily compete with the Standard Model and possibly even
dominate $M_{12}$. If the non-standard contributions to $M_{12}$ are
unrelated to the CKM mechanism of the { three-generation} Standard Model,
they will affect the mixing phase 
\begin{eqnarray} 
\phi_M &=& \arg M_{12} . \no
\end{eqnarray} 
With our convention \eq{defcp} the Standard Model prediction is
$\phi_M=\arg (V_{tb} V_{ts}^*)^2$.

The mass eigenstates at time $t=0$, $\ket{B_L}$ and $\ket{B_H}$, are
linear combinations of $\ket{B_s}$ and $\ket{\ov{B}{}_s}$:
\begin{eqnarray}
\mbox{lighter eigenstate:} \quad \ket{B_L} &=& 
        p \ket{B_s} + q \ket{\ov{B}{}_s} \nn
\mbox{heavier eigenstate:} \quad \ket{B_H} &=& 
        p \ket{B_s} - q \ket{\ov{B}{}_s}, 
\qquad \qquad \mbox{with $|p|^2+|q|^2=1$}.       
\label{defpq}
\end{eqnarray}
We denote the masses and widths of the two eigenstates with $M_{L,H}$
and $\Gamma_{L,H}$ and define 
\begin{eqnarray}
\Gamma \; = \; \frac{1}{\tau_{B_s}} &=& \frac{\Gamma_H+ \Gamma_L}{2}, 
\qquad\quad  
\dm \; = \; M_H - M_L ,  
\qquad\quad 
\dg \; = \; \Gamma_L - \Gamma_H 
 . \label{defdm}
\end{eqnarray}
While $\dm >0$ by definition, $\dg$ can have either sign. Our sign
convention is such that $\dg >0$ in the Standard Model.  
By examining the eigenvalue problem of $M-i\Gamma/2$ we find that
the experimental information  $\dm \gg \Gamma$ model-independently 
implies $|\Gamma_{12}| \ll |M_{12}|$. By expanding the eigenvalues and
$q/p$ in $\Gamma_{12}/M_{12}$, we find 
\begin{eqnarray}
\dm &=& 2 |M_{12}|, \qquad \dg \; = \; 
        2 \lt| \Gamma_{12} \rt| \cos \phi \qquad 
\mbox{and} \qquad \frac{q}{p} \; = \; - e^{-i \phi_M} 
        \lt[ 1 - \frac{a}{2} \rt] 
 . \label{dmdg}
\end{eqnarray}
Here the phase $\phi$ is defined as 
\begin{eqnarray} 
\frac{M_{12}}{\Gamma_{12}} = - \lt| \frac{M_{12}}{\Gamma_{12}}  \rt|\, 
  e^{i \phi} \label{defphi}. 
\end{eqnarray}  
In \eq{dmdg} we have kept a correction in the small parameter
\begin{eqnarray}
a &=&  \lt| \frac{\Gamma_{12}}{M_{12}}  \rt| \sin \phi , \label{defa} 
\end{eqnarray}
but neglected all terms of order $\Gamma_{12}^2/M_{12}^2$ and do so
throughout this paper. Since $a$ can hardly exceed $0.01$ we will
likewise set it to zero in our studies of $B_s$ decays into CP
eigenstates and only briefly discuss a non-zero $a$ in 
sect.~\ref{sec:cpmix}.

The phase $\phi$ is physical and convention-independent; if $\phi = 0 $, CP
violation in mixing vanishes.  In the Standard Model $\phi = \phi_M -
\arg (-\Gamma_{12})$ is tiny, of order 1\%. This is caused by two
effects: first, $\Gamma_{12}$ is dominated by the decay $b\to c \ov{c}
s$ and $(V_{cb} V_{cs}^*)^2$ is close to the \bbms\ phase $\arg
(V_{tb} V_{ts}^*)^2$. Second, the small correction to $\arg
(-\Gamma_{12})$ involving $V_{ub} V_{us}^*$ is further suppressed by a
factor of $m_c^2/m_b^2$. { In the search for a sizeable new physics
contribution to $\phi$ these doubly Cabibbo-suppressed terms
proportional to $V_{ub} V_{us}^*$ can safely be neglected, as we do
throughout this paper.}

% This is not surprising, because $M_{12}$ and
% $\Gamma_{12}$ are dominated by two quark families only (the second and
% third) and { unitarity of the corresponding $2 \times 2$ CKM
% submatrix holds to a few percent. There is no CP violation in a
% two-family Standard Model}.

For a $B_s$ decay into { some final state} $f$, we introduce the \dbo\
matrix elements
\begin{eqnarray}
   A_f = \langle f \ket{B_s} 
 && \qquad\quad \mbox{and} \qquad\quad
   \ov{A}_f = \langle f \ket{\ov{B}_s} .\no
\end{eqnarray} 
The key quantity for CP violation reads  
\begin{eqnarray}
\lambda_f & = & \frac{q}{p}\, \frac{\ov{A}_f}{A_f}
    . \label{deflaf}
\end{eqnarray}
The time evolution formulae and the expressions for the CP asymmetries
in the forthcoming sections can be conveniently expressed in terms of 
\begin{eqnarray}
\adi = \frac{1- \lt| \lambda_f \rt|^2}{1+ \lt| \lambda_f 
        \rt|^2} , && \qquad\quad
\ami = - \frac{2\, \imag \lambda_f}{1+ \lt| \lambda_f \rt|^2} 
   \qquad\quad \mbox{and} \qquad\quad
\adg = - \frac{2\, \real \lambda_f}{1+ \lt| \lambda_f   \rt|^2} 
        \, . \label{defacp} 
\end{eqnarray}
If $f$ is a CP eigenstate, $CP\ket{f}=\pm \ket{f}$, then $\adi \neq 0$
or $\ami \neq 0$ signals CP violation: a non-vanishing $\adi$ implies $
|A_f| \neq |\ov{A}_f|$, meaning direct CP violation; $\ami$ measures
mixing-induced CP violation in the interference of $B_s \to f$ and
$\ov{B}_s \to f$. The third quantity, $\adg$, plays a role, if \dg\ is
sizeable.  The three quantities obey the relation
\begin{eqnarray}
\lt| \adi \rt|^2 + \lt| \ami \rt|^2 + \lt| \adg \rt|^2 &=& 1    . \no
\end{eqnarray}

The time-dependent decay rate \gtf\ of an initially tagged $B_s$
into some final state $f$ is defined as 
\begin{eqnarray} 
\gtf &=& \frac{1}{N_B}\,
        \frac{d N(B_s (t) \rightarrow f)}{d t} \, .
\label{defgtf}
\end{eqnarray}
Here $B_s(t)$ represents a meson { at proper time} $t$ tagged as a
$B_s$ at $t=0$;  $d N(B_s(t) \rightarrow f)$ denotes the number of
decays of $B_s(t)$ into the final state $f$ occurring within the time
interval $[t,t+d t]$; $N_B$ is the total number of $B_s$'s produced at
time $t=0$.  An analogous definition holds for \gbtf.  By solving the
Schr\"odinger equation \eq{schr} using \eq{dmdg}, we can find these
decay rates \cite{dr}:
\begin{eqnarray} 
\!\!\! \gtf &=&  {\cal N}_f \, | A_f |^2 \, \frac{1 + \lt| \lambda_f \rt|^2}{2}
        \, e^{-\Gamma t}  \no \\*
&& \times \lt[ \cosh \frac{\dg \, t}{2} \, + \,  
   \adi \, \cos ( \dm \, t )  
  + \adg \, \sinh \frac{\dg \, t}{2} 
  + \ami \, \sin \lt( \dm \, t \rt) \rt] ,  
\label{gtfres} \\
\!\!\! \gbtf &=& {\cal N}_f \, | A_f |^2 \,  
        \frac{1 + \lt| \lambda_f \rt|^2}{2}\, ( 1 + a ) \,
  e^{-\Gamma t}  \no \\* 
&& \times  \lt[  
    \cosh \frac{\dg \, t}{2}  
  - \adi \, \cos ( \dm \, t ) 
        + \adg \, \sinh \frac{\dg \, t}{2} 
        - \ami \, \sin ( \dm \, t ) \rt]  .
   \label{gbtfres}
\end{eqnarray}
Here ${\cal N}_f$ is a time-independent normalization factor. 

\begin{ntable}
\centerline{
\begin{tabular}{@{}r|p{0.49\textwidth}|p{0.22\textwidth}}
\hline
Quark decay & Hadronic decay & Remarks \\\hline\hline  
\phantom{\Large X}
$\ov{b} \rightarrow \ov{c} c \ov{s}$ 
& $B_s \rightarrow \psi \phi$ &  \\
& $B_s \rightarrow \psi K^{(*)} \ov{K}{}^{(*)}$ & \\
& $B_s \rightarrow \psi \phi \phi$ & \\
& $B_s \rightarrow \psi \eta $\\ 
& $B_s \rightarrow \psi \eta^\prime$ & \\ 
& $B_s \rightarrow \psi f_0 $ & CP-odd final state\\
& $B_s \rightarrow \chi_{c0} \phi $ & CP-odd final state\\
& $B_s \rightarrow D_s^{(*)}{}^+ D_s^{(*)}{}^-$ & $D_s^+ D_s^-$ is CP-even\\
& $B_s \rightarrow D^{(*)}{}^+ D^{(*)}{}^-$  or  
        $D^{(*)}{}^0 \ov{D}{}^{(*)}{}^0$ &
        non-spectator decays, 
        $D \ov{D}$ is CP-even \\\hline 
\phantom{\Large X}
$\ov{b} \rightarrow \ov{c} u \ov{d}$ 
& $B_s \rightarrow K_S \ov{D}{}^{(*)0}$
  [$\rightarrow  \phi K_S$, $\rho^0 K_S$, $  K \ov{K}$ 
        or $\pi^+ \pi^- $] &  \\\hline
\end{tabular}
}
\caption{Some { CKM-favoured} $B_s$ decay modes into CP-specific final
states. Here, $\psi$ represents $J/\psi$ or $\psi(2S)$.  Decays into two
vector particles or into three-body final states with one or more
vector particles require an angular analysis to separate the CP-even
from the CP-odd component. The final states $D_s^{\pm} D_s^{*}{}^{\mp}$
are dominantly CP-even \cite{ayopr} (see sect.~\ref{sec:untagged}).
}\label{tab}
\end{ntable}
{ A promising}
testing ground for new physics contributions to $\phi_M$ are
decays into CP eigenstates triggered by the quark decay $b\to
c\ov{c}s$. Table~\ref{tab} summarizes such CP-specific $B_s$ decay
modes.  To estimate the size of the small Standard Model predictions
consider first the decay amplitudes \cite{RF-psiK}:
\begin{eqnarray}
A_f,\ov{A}_f & \propto& 
 \left[1+\left(\frac{\lambda^2}{1-\lambda^2}\right)
        \, a_{ \rm p} \, e^{i\theta} \, e^{\pm i\gamma}\right].
 \label{ampl-ratio}
\end{eqnarray}
Hence the weak phase factor $e^{i\gamma}$, which is associated with
the quantity $a_{ \rm p} e^{i\theta}$, is strongly
Cabibbo-suppressed by two powers of the Wolfenstein parameter
$\lambda\simeq |V_{us}|\simeq 0.22$ \cite{wolf}.  The ``penguin
parameter'' $a_{ \rm p} e^{i\theta}$ measures -- sloppily speaking
-- the ratio of penguin- to tree-diagram-like topologies and is
loop-suppressed.  Since new-physics contributions to these decay
amplitudes have to compete with a tree diagram, they { are not
  expected to play a significant role}. A detailed discussion for a
left--right-symmetric model can be found in \cite{bfNP}.  Since we are
interested in large ``smoking gun'' new physics effects {in \bbms}, we
account for the Standard Model contributions within the leading order
of $\lambda$ and set $|\ov{A}_f|=|A_f|$, neglecting  direct CP
  violation. With the weak phase $\phi_{ c\ov{c} s}=\arg (V_{cb}
V_{cs}^*)$ one then finds
\begin{eqnarray}
\frac{\ov{A}_f}{A_f} &=& - \eta_f e^{2 i \phi_{ c\ov{c} s}} . \label{afaf}
\end{eqnarray}
Here $\eta_f$ denotes the CP parity of $f$: $CP \ket{f}= \eta_f \ket{f}$.
In Table~\ref{tab} we also included decay modes driven by the
quark level decay $b\to c\ov{u}d$.  The weak phase of these modes
involves the phases of the $K$ and $D$ decay amplitudes into CP
eigenstates. The phases combine to $\arg (V_{cb} V_{ud}^*) +
\arg (V_{ud} V_{us}^*) + \arg (V_{us} V_{cs}^*) = \arg (V_{cb}
V_{cs}^*)$, i.e.\ the same result as for $b\to c\ov{c}s$. With
\eq{dmdg} and \eq{afaf} $\lambda_f$ reads
\begin{eqnarray}
\lambda_f & = & \frac{q}{p}\, \frac{\ov{A}_f}{A_f}
  \; = \; \eta_f \, e^{- i \phi }  
    . \label{laf}
\end{eqnarray}
Here we have identified the phase $\arg (\eta_f \lambda_f) =
\phi_M - 2 \phi_{c\ov{c} s}$ with the phase $\phi$ defined in
\eq{defphi}. This is possible, because $\arg (-\Gamma_{12})=2 \phi_{
c\ov{c} s} +O (\lambda^2)$ and we neglect the Cabibbo-suppressed
contributions. The Standard Model contribution to $\phi=\phi_{\rm SM} +
\phi_{\rm NP}$ equals $\phi_{\rm SM}=-2 \eta \lambda^2 $. Here $\eta$ is the
Wolfenstein parameter measuring the height of the unitarity
triangle. Since our focus is a sizeable new physics contribution
$\phi_{\rm NP}$, we can safely neglect $\phi_{\rm SM}$ and identify $\phi$
with $\phi_{\rm NP}$ in the following. That is, we neglect terms of order 
$\lambda^2$ and higher.  Using \eq{laf} the quantities in
\eq{defacp} simplify to
\begin{eqnarray} 
\adi = 0 
        , && \qquad\quad
\ami = \eta_f \sin \phi 
   \qquad\quad \mbox{and} \qquad\quad
\adg = - \eta_f \cos \phi
        . \label{acp2} 
\end{eqnarray}
{ The corrections to \eq{acp2} from penguin effects} can be found
in \cite{RF-psiK}. 
We next specify to the PDG phase convention for the CKM matrix
\cite{pdg}, in which $\arg (V_{cb} V_{cs}^*) = {\cal O}
(\lambda^6)$. Then we can set $\phi_{c\ov{c} s}$ to zero and identify
\begin{eqnarray}
\phi_M &=& \phi .  \no  
\end{eqnarray}
With this convention the mass eigenstates { can be expressed as}
\begin{eqnarray}
\ket{B_L} & = & \phantom{- \,}
                \frac{1+e^{i\phi}}{2} \, 
                  \ket{B_s^{\textrm{\scriptsize even}}} \, { -} \, 
                  \frac{1-e^{i\phi}}{2} \, 
                  \ket{B_s^{\textrm{\scriptsize odd}}} \, + \, 
        {\cal O} (a) , \nn
\ket{B_H} & = & { -} \, \frac{1-e^{i\phi}}{2} \, 
                  \ket{B_s^{\textrm{\scriptsize even}}} \, + \, 
                  \frac{1+e^{i\phi}}{2} \, 
                  \ket{B_s^{\textrm{\scriptsize odd}}} \, + \, 
        {\cal O} (a) \, . 
        \label{rot}
\end{eqnarray} 
{ Whenever we use $B_s^{\textrm{\scriptsize even}}$ and
  $B_s^{\textrm{\scriptsize odd}}$ we implicitly refer to this phase
  convention.  If formulae involving $B_s^{\textrm{\scriptsize even}}$
  and $B_s^{\textrm{\scriptsize odd}}$ are used to constrain models
  with an extended quark sector, the phase convention used for the
  enlarged CKM matrix must likewise be chosen such that $\arg (V_{cb}
  V_{cs}^*) \simeq 0$.}  

% We close this section with a word of caution: experimental
% considerations for the measurement of \dg\ have also included the
% decay mode $B_s \to K^+ K^-$ because of its nice signature.  { This
% decay mode only involves the same weak phase as the modes listed in
% Table~\ref{tab}, if both the tree amplitude is negligible and the
% penguin amplitude is dominated by the Standard Model
% contribution. While the assumption of penguin dominance is even
% doubtful in the Standard Model, most extensions of the Standard Model
% which affect \bbms\ likewise modify the penguin amplitude triggering
% $B_s \to K^+ K^-$.  One can certainly expect a sizeably nonzero $\adi$
% in this mode.  Therefore} the three expressions in \eq{acp2} do not
% hold and the formulae in this paper cannot be used for this decay
% mode. The same remark of course applies to other penguin-induced decay
% modes like $B_s \to K_S K_S $ or $B_s \to \phi \phi $.
%
%
\section{Untagged Studies}\label{sec:untagged}
\subsection{Time Evolution}
Whereas the width difference \dg\ is negligibly small in the $B_d$
system, it can be sizeable for $B_s$ mesons. This has the consequence
that the untagged $B_s$ data sample bears information on CP violation
\cite{dun}. Further the width difference itself is sensitive to the
\bbms\ phase $\phi$ \cite{g}, as we can see from \eq{dmdg}. 

When $B_s$'s and $\ov{B}_s$'s are produced in equal numbers, the 
untagged decay rate for the decay $\Bsun \rightarrow f$ reads
\begin{eqnarray}
\guntf &=& \gtf + \gbtf        \no \\*
&=&  {\cal N}_f
        \lt[ \,  e^{-\Gamma_L t} \lt| \langle f \ket{B_L} \rt|^2 + 
        e^{-\Gamma_H t} \lt| \langle f \ket{B_H} \rt|^2  \rt] +  
        {\cal O} ( a ) . \label{twoex} \\*
&=& {\cal N}_f \lt| A_f \rt|^2  
        \lt[ 1 + |\lambda_f|^2  \rt] \, e^{- \Gamma t}  \lt\{ 
        \cosh \frac{\dg \,  t}{2} +  \, 
        \sinh \frac{\dg \,  t}{2} \, \adg
        \rt\} 
        + {\cal O} ( a ) \label{guntf} .
\end{eqnarray}
Here the second expression is simply obtained by adding \eq{gtfres} and 
\eq{gbtfres}. In \eq{twoex} the same result is expressed in terms of 
the mass eigenstates and nicely exhibits how the decay is governed by
two exponentials. Using \eq{defgtf} we can relate the overall
normalization to the branching ratio:
\begin{eqnarray} 
\brunt{f} \, &=& 
% \frac{1}{2} Br \, \big( B_s (t) \rightarrow f \, \big)\,        
%     +   \frac{1}{2} Br \, \big( \ov{B}_s (t) \rightarrow f \, \big)
% \; = \;   
\frac{1}{2} \int_0^\infty \! dt \,\, \guntf\
        \label{br} \\*
&=&  \frac{{\cal N}_f}{2}   \, | A_f |^2 \, \lt[ 1+ | \lambda_f |^2 \rt]  
        \frac{\Gamma + \adg \,
        \dg/2 }{\Gamma^2-(\dg/2)^2} 
        + {\cal O} ( a ) .
        \label{untnorm} 
\end{eqnarray}
Conforming with \cite{pdg} we have normalized the event counting to
$N_B+N_{\ov{B}}=2N_B$, 
so that $\brunt{all} =1$.
Using \eq{untnorm} we rewrite \eq{guntf} as
\begin{eqnarray}
\guntf & = &
   2 \, \brunt{f} \,\, 
        \frac{\Gamma^2-(\dg/2)^2}{\Gamma + \adg \, \dg/2} 
        \,\, e^{- \Gamma t} \, \lt[ 
        \cosh \frac{\dg \,  t}{2} +  \, 
        \sinh \frac{\dg \,  t}{2} \, \adg
        \rt] 
        + {\cal O} ( a ) 
. \label{guntf2} 
\end{eqnarray}
Now \eq{guntf2} is our master equation for the time evolution of the
decay of an untagged $B_s$ sample. If $\Gamma=1/\tau_{B_s}$ is known,
one could perform a two-parameter fit of the decay distribution to
\eq{guntf2} and determine $\dg$ and $\adg$. { The latter determines
$\phi$ through \eq{acp2}, if $f$ is a CP eigenstate from a
CKM-favoured decay.} In practice, however, most data come from short
times with $\dg \, t \ll 1$, and one is only sensitive to the product
$\dg \cdot \adg$:
\begin{eqnarray}
\guntf & = & 
2\, \brunt{f} \,\,
        \Gamma 
        \, e^{- \Gamma t} \, \lt[ 
        1 +  \frac{\dg}{2} \, \adg \, 
        \lt( t - \frac{1}{ \Gamma } \rt) 
        \rt] 
        + {\cal O} \lt(\lt( \dg\, t \rt)^2 \rt) . \label{guntf3} 
\end{eqnarray} 
We return to this point in sect.~\ref{sec:detdg}.

\boldmath
\subsection{The Width Difference \dg\ and Branching Ratios}\label{sec:wd}
\unboldmath
{ The mass matrix $M_{12}$ and the decay matrix $\Gamma_{12}$
provide three rephasing invariant quantities: $|M_{12}|$, 
$|\Gamma_{12}|$ and the relative phase $\phi$. In \eq{dmdg} we have
related the two observables \dm\ and \dg\ to $|M_{12}|$, 
$|\Gamma_{12}|$ and $\phi$. Interestingly, it is possible to find 
a third observable, which determines $|\Gamma_{12}|$ and 
thus encodes additional information. 
We define
\begin{eqnarray}
\dg_{\rm CP} &\equiv& 2 |\Gamma_{12}| \; = \; 
2\, \sum_{f  \in X_{c\ov{c}}} \,  \lt[
\Gamma ( B_s \to f_{\rm CP+} ) \, - \,
        \Gamma ( B_s \to f_{\rm CP-}) \rt]
        . 
        \label{dgcp}
\end{eqnarray}
Here $X_{c\ov {c}}$ represents the final states containing a
  $(c,\ov{c})$ pair, which constitute the dominant contribution to
  $\dg_{\rm CP}$ stemming from the decay $b \to c\ov{c} s$.}  In
  \eq{dgcp} we have decomposed any final state $f$ into its CP-even
  and CP-odd component, $\ket{f}=\ket{f_{\rm CP+}} + \ket{f_{\rm
  CP-}}$ and defined
\begin{eqnarray}
\Gamma ( B_s \to f_{\rm CP\pm} ) &=& 
        {\cal N}_f \, | \bra{f_{\rm CP\pm}} B_s \rangle |^2 
  \; = \; 
     \frac{ | \bra{f_{\rm CP\pm}} B_s \rangle |^2}{ | \bra{f} B_s \rangle
     |^2} \, 
        \Gamma ( B_s \to f ) 
. \no 
\end{eqnarray}
${\cal N}_f$ is the usual normalization factor originating from
the phase-space integration.  In order to prove the second equality in
\eq{dgcp} we start from the definition of $\Gamma_{12}$:
\begin{eqnarray}
\Gamma_{12} &=& \sum_f {\cal N}_f \, \bra{B_s} f \rangle 
                \bra{f} \ov{B}_s \rangle 
        \; = \; \frac{1}{2} \sum_f {\cal N}_f  
        \lt[ \, \bra{B_s} f \rangle \bra{f} \ov{B}_s \rangle + 
        \bra{B_s} \ov{f} \rangle \bra{{ \ov{f}}} \ov{B}_s \rangle \, \rt]
 .\label{defg12} 
\end{eqnarray}
In the second equation we have paired the final state $\ket{f}$ with its CP
conjugate $\ket{\ov{f}}=-CP\ket{f}$.
In the next step we 
trade $f$ for $f_{\rm CP+}$ and $f_{\rm CP-}$ and use the CP transformation 
\begin{eqnarray}
 \bra{f_{\rm CP\pm}} \ov{B}_s \rangle  &=& 
        \mp \, e^{2 i \phi_{c\ov{c} s}} \, \bra{f_{\rm CP\pm}} B_s \rangle 
  ,  \no 
\end{eqnarray}
where $\phi_{c\ov{c}s} = \arg (V_{cb} V_{cs}^*) $ is the phase of the $b\to c
\ov{c} s$ decay amplitude, which dominates $\Gamma_{12}$. Then
\eq{defg12} becomes 
\begin{eqnarray}
- \,  e^{-2 i \phi_{c\ov{c}s}} \, \Gamma_{12} &=& 
        \sum_{f\in X_{c\ov{c}}} {\cal N}_f \,
        \lt[ | \bra{f_{\rm CP+}} B_s \rangle |^2 - 
             | \bra{f_{\rm CP-}} B_s \rangle |^2 \rt] \nn
& = & \sum_{f\in X_{c\ov{c}}}
\lt[ \Gamma ( B_s \to f_{\rm CP+} ) \, - \,
        \Gamma ( B_s \to f_{\rm CP-}) \rt] .
\label{g12oe}
\end{eqnarray}
Interference terms involving both $ \bra{f_{\rm CP+}} B_s \rangle$ and
$\bra{f_{\rm CP-}} B_s \rangle $ drop out when summing the two terms
$\bra{B_s} f \rangle \bra{f} \ov{B}_s \rangle$ and $\bra{B_s} \ov{f}
\rangle \bra{{ \ov{f}}} \ov{B}_s \rangle $.  In \eq{g12oe} both sides
of the equation are rephasing-invariant.  An explicit calculation of
$\Gamma_{12}$ reveals that the overall sign of the LHS of \eq{g12oe}
is positive, which completes the proof of \eq{dgcp}.

Loosely speaking, $\dg_{\rm CP}$ is measured by counting the CP-even and
CP-odd double-charm final states in $B_s$ decays. We specify this
statement in the following and relate $\dg_{\rm CP}$ to measured
observables in sect.~\ref{sec:both}. Our formulae become more
transparent if we adopt the standard phase convention with $\arg
(V_{cb} V_{cs}^*)\simeq 0$ and use the CP-eigenstates defined in
\eq{cpe}. With $\ket{B_s}= (\ket{B_s^{\textrm{\scriptsize
even}}}+\ket{B_s^{\textrm{\scriptsize odd}}})/\sqrt{2}$ one easily
finds from \eq{g12oe}:
\begin{eqnarray}
\dg_{\rm CP} & = & 2 |\Gamma_{12}| \; = \; 
\Gamma \lt( B_s^{\textrm{\scriptsize even}} \rt) - 
        \Gamma ( B_s^{\textrm{\scriptsize odd}} ) .
        \label{dgcp2}
\end{eqnarray}
Here the RHS refers to the total widths of the CP-even and CP-odd
$B_s$ eigenstates.  We stress that the possibility to relate
$|\Gamma_{12}|$ to a measurable quantity in \eq{dgcp} crucially
depends on the fact that $\Gamma_{12}$ is dominated by a single weak
phase. For instance, the final state $K^+ K^-$ is triggered by $b \to u
\ov{u} s$ and involves a weak phase different from $b \to c\ov{c} s$. 
{Although} $K^+ K^-$ is CP-even, the decay $B_s^{\textrm{\scriptsize
odd}}\to K^+ K^- $ is possible. An
inclusion of { such} CKM-suppressed modes into \eq{g12oe} would add
interference terms that spoil the relation to measured quantities.
{ The omission of these contributions to
$\Gamma_{12}$ induces a theoretical uncertainty of order 5\% on \eq{dgcp2}.}

In the Standard Model the mass eigenstates in \eq{rot} coincide with
the CP eigenstates (with $B_L=B_s^{\textrm{\scriptsize even}}$) and
$\dg_{\rm SM}=\dg_{\rm CP}$. 
The effect of a non-zero \bbms\ phase $\phi$ reduces
$\dg$: 
\begin{eqnarray} 
\dg &=& \dg_{\rm CP} \cos \phi ,\label{dgcph} 
\end{eqnarray} 
while $\dg_{\rm CP}=2 |\Gamma_{12}|$ is not sensitive to new
physics. From the calculated $\Gamma_{12}$ we can predict to which
extent $\Gamma \lt( B_s^{\textrm{\scriptsize even}} \rt)$ exceeds 
$\Gamma \lt( B_s^{\textrm{\scriptsize odd}} \rt)$ and this result does
not change with the presence of a non-zero $\phi$.

The theoretical prediction for $\dg_{\rm CP}$ is known to next-to-leading
order in both $\Lambda_{\rm QCD}/m_b$ \cite{bbd} and the QCD coupling
$\alpha_s$ \cite{bbgln}. It reads  
\begin{eqnarray} 
\frac{\dg_{\rm CP}}{\Gamma} &=& 
        \left( \frac{f_{B_s}}{245~{\rm MeV}} \right)^2 \,
        \left[ \, (0.234\pm 0.035)\, B_S - 0.080 \pm 0.020 \, \right] .
\label{dgnum} 
\end{eqnarray} 
Here the coefficient of $B_S$ has been updated to
$m_b(m_b)+m_s(m_b)=4.3\,$GeV (in the $\ov{\rm MS}$ scheme) and
$f_{B_s}$ is the $B_s$ meson decay constant. Recently the
KEK--Hiroshima group succeeded in calculating $f_{B_s}$ in an
unquenched lattice QCD calculation with two dynamical fermions
\cite{fbs}. The result is $f_{B_s}=(245\pm 30)\,$MeV. $B_S$
parametrizes the relevant hadronic matrix element, with $B_S=1$ in the
vacuum saturation approximation. A recent quenched lattice calculation
has yielded $B_S=0.87 \pm 0.09$ \cite{hioy} for the $\ov{\rm MS}$
scheme. A similar result has been found in \cite{bmrglm}. This
analysis, however, calculates \dg\ after normalizing \eq{dgnum} to the
measured mass difference in the \bbd\ system.  This method involves $|V_{td}|$,
which is obtained from a global CKM fit and thereby relies on the
Standard Model. Since the target of our analysis is new physics, we
cannot use the numerical prediction for \dg\ of \cite{bmrglm}.  At
present, studies of $B_S$ are a new topic in lattice calculations and
we can expect substantial improvements within the next few years. With
these numbers one finds from \eq{dgnum}:
\begin{eqnarray} 
\frac{\dg_{\rm CP}}{\Gamma} &=& 0.12\pm 0.06 
        . \label{dgnum2}
\end{eqnarray}
Here we have conservatively added the errors from the two lattice quantities 
linearly.  

Since $\dg_{\rm CP}$ is unaffected by new physics and $\dg_{\rm CP} > 0$,
several facts hold beyond the Standard Model: i) There are more
CP-even than CP-odd final states in $B_s$ decays. ii) The
shorter-lived mass eigenstate is always the one with the larger
CP-even component in \eq{rot}. Its branching ratio into a CP-even
final state $f_{\rm CP+}$ exceeds the branching ratio of the longer-lived
mass eigenstate into $f_{\rm CP+}$, 
if { the weak phase of the decay amplitude is close to $ \arg
V_{cb} V_{cs}^*$.} 
For $ \cos \phi > 0$ $B_L$ has a shorter lifetime than $B_H$, while for $
\cos \phi < 0$ the situation is the opposite \cite{g}. iii) Measurements
based on the comparison of \emph{branching ratios}\ into CP-specific
final states determine $\dg_{\rm CP}$ rather than $\dg$. Such an analysis
has recently been performed by the ALEPH collaboration \cite{aleph}.
ALEPH has measured
\begin{eqnarray}
% 2\, Br(\Bsun \to D_s^{(*)}{}^+ D_s^{(*)}{}^-) & = & 
 2\, \brunt{D_s^{(*)}{}^+ D_s^{(*)}{}^-} & = & 
        0.26 \epm{0.30}{0.15} \label{alexp}
\end{eqnarray}
and related it to $\dg_{\rm CP}$. For this the following theoretical input has
been used \cite{ayopr}: 
\begin{itemize} 
\item[i)] In the heavy quark limit $m_c \to \infty $ and 
 neglecting certain terms of order 
 $1/N_c$ (where $N_c=3$ is the number of colours)
 the decay $B_s^{\textrm{\scriptsize odd}} \to D_s^\pm D_s^{*}{}^\mp$
 is forbidden. Hence in this limit the final state in 
 $\Bsun \to  D_s^\pm D_s^{*}{}^\mp$ is CP-even. Further in 
 $\Bsun \to D_s^{*}{}^+ D_s^{*}{}^-$ the final state is in an 
 S-wave. 
\item[ii)] In the Shifman--Voloshin (SV) limit $m_c \to \infty $ with
 $m_b - 2 m_c \to 0$ \cite{sv}, $\dg_{\rm CP}$ is saturated by $\Gamma ( \Bsun \to
 D_s^{(*)}{}^+ D_s^{(*)}{}^- )$.
 With i) this implies that in the considered limit the width of 
 $B_s^{\textrm{\scriptsize odd}}$ vanishes.
 For $N_c \to \infty$ and in the SV limit, ${ 2 \Gamma ( \Bsun \to
 D_s^{(*)}{}^+ D_s^{(*)}{}^- )}$ further equals the parton model result for
 $\dg_{\rm CP}$ (quark-hadron duality).  
\end{itemize}
Identifying  
$\Gamma ( B_s^{\textrm{\scriptsize even}} \to  D_s^{(*)}{}^+ D_s^{(*)}{}^- ) 
\simeq \dg_{\rm CP}$ and
$\Gamma ( B_s^{\textrm{\scriptsize odd}} \to  D_s^{(*)}{}^+ D_s^{(*)}{}^- ) 
\simeq 0$ we find:
\begin{eqnarray}
\!\!\!\!
2\, \brunt{ D_s^{(*)}{}^+ D_s^{(*)}{}^-} & \simeq & 
        \dg_{\rm CP} \, \lt[ \, \frac{1+\cos \phi}{2\, \Gamma_L}  + 
                            \frac{1-\cos \phi}{2\, \Gamma_H} \, \rt] 
         \; = \;  \frac{\dg_{\rm CP}}{\Gamma} \lt[ 
  1 + {\cal O} \lt( \frac{\dg}{\Gamma} \rt) \rt] . \label{bdg}
\end{eqnarray}
Thus the measurement in \eq{alexp} is compatible with the theoretical
prediction in \eq{dgnum2}. For $\phi=0$, the expression used in
Ref.~\cite{aleph}, in which the Standard Model scenario has been
considered, is recovered. The term in square brackets accounts for the
fact that in general the CP-even eigenstate
$\ket{B_s^{\textrm{\scriptsize even}}}$ is a superposition of
$\ket{B_L}$ and $\ket{B_H}$. It is straightforward to obtain \eq{bdg}:
inserting \eq{rot} into \eq{twoex} expresses \guntf\ in terms of
$\Gamma ( B_s^{\textrm{\scriptsize even}} \to f )$ and $\Gamma (
B_s^{\textrm{\scriptsize odd}} \to f )$. After integrating over time
the coefficient of $\Gamma ( B_s^{\textrm{\scriptsize even}} \to f )$
is just the term in square brackets in \eq{bdg}.

When using \eq{bdg} one should be aware that the corrections to the
limits i) and ii) adopted in \cite{ayopr} can be numerically
sizeable. For instance, in the SV limit there are no multibody final
states like $D_s^{(*)} \ov{D} X_s$, which can modify \eq{bdg}. As
serious would be the presence of a sizeable CP-odd component of the
$D_s^{(*)}{}^+ D_s^{(*)}{}^- $ final state, since it would be added
with the wrong sign to $\dg_{\rm CP}$ in \eq{bdg}. A method to control
the corrections to the SV limit experimentally is proposed in
sect.~\ref{sec:both}.  We further verify from \eq{bdg} that the
measurement of $\brunt{D_s^{(*)}{}^+ D_s^{(*)}{}^-}$ determines
$\dg_{\rm CP}$.  Its sensitivity to the new physics phase $\phi$ is
suppressed by another factor of $\dg/\Gamma$ and is irrelevant in view of
the theoretical uncertainties.

\boldmath
\subsection{Determination of \dg\ and $|\cos \phi|$}\label{sec:detdg}
\unboldmath
There are two generic ways to obtain information on \dg :
\begin{itemize}
\item[i)] The measurement of the $B_s$ lifetime in two decay modes 
        $\Bsun \to f_1$ and  $\Bsun \to f_2$ with\\
        $\adg (f_1) \neq \adg (f_2)$. 
\item[ii)] The fit of the decay distribution of 
        $\Bsun \to f$ to the two-exponential
        formula in \eq{guntf2}.  
\end{itemize}
As first observed in \cite{g}, the two methods are differently affected
by a new physics phase $\phi \neq 0$. Thus by combining the results of
methods i) and ii) one can determine $\phi$.  In this section we
consider two classes of decays:
\begin{itemize}
\item flavour-specific decays, which are characterized by
        $\ov{A}_f=0$ { implying} $\adg=0$. 
        Examples are $B_s \to D_s^- \pi^+$ and 
        $B_s \to X \ell^+ \nu_{\ell} $,  
\item the CP-specific decays of Table~\ref{tab}, with 
        $\adg = - \eta_f \cos \phi$.  
\end{itemize}
In both cases the time evolution of the untagged sample in \eq{guntf2}
is not sensitive to the sign of $\dg$ {(or, equivalently, of $\cos
  \phi$)}. For the CP-specific decays of Table~\ref{tab} this can be
seen by noticing that
\begin{eqnarray}
\adg\, \sinh \frac{\dg \, t}{2} & = & 
        - \, \eta_f \, |\cos \phi| \, \sinh \frac{|\dg| \, t}{2} .\no
\end{eqnarray}
Here we have used the fact that $\dg $ and $\cos \phi $ always have the same
sign, because $\dg_{\rm CP}>0$.  Hence the untagged studies discussed here
in sect.~\ref{sec:detdg} can only determine $|\cos \phi|$ and
therefore lead to a four-fold ambiguity in $\phi$.  The sign ambiguity
in $\cos \phi$ reflects the fact that from the untagged time
evolution in \eq{guntf2} one cannot distinguish, whether the
heavier or the lighter eigenstate has the shorter lifetime 
(however, see sect.~5).

{ In order to experimentally establish a non-zero \dg\ from the
time evolution in \eq{guntf2} one needs sufficient statistics to
resolve the deviation from a single-exponential decay law, see 
\eq{guntf3}.  As long as we are only sensitive to terms linear in
$\dg\, t$ and $\dg/\Gamma$, we can only determine $\adg\, \dg$ from
\eq{guntf3}. $\adg\, \dg$ vanishes for flavour-specific decays and
equals $-\eta_f \dg\, \cos \phi$ for CP-specific final states. Hence
from the time evolution alone one can only determine $\dg\, \cos \phi
$ in the first experimental stage.  This determination is discussed in
sect.~\ref{sec:dgcos}. Once the statistical accuracy is high enough to
resolve terms of order $(\dg)^2$, one can determine both $|\dg|$ and 
$|\cos \phi|$. Fortunately, the additional information from branching
ratios can be used to find $|\dg|$ and $|\cos \phi|$ without resolving 
quadratic terms in \dg. The determination of $|\dg|$ and $|\cos \phi|$
is discussed in sect.~\ref{sec:both}.}  
  
\boldmath
\subsubsection{Determination of $\Gamma$ and $\dg \cos \phi$}\label{sec:dgcos}
\unboldmath
Lifetimes are conventionally measured by fitting the decay
distribution to a single exponential. Consider a decay which is
governed by two exponentials,
\begin{eqnarray}
%\!\!\!\!\!\!
{ 
\frac{\guntf +\guntfb }{2} }
&=& A \, e^{-\Gamma_L t} \, + \, B \, e^{-\Gamma_H t} \nn
       & = & e^{-\Gamma t} \lt[ 
        \lt( A+B \rt) \cosh \frac{\dg t}{2} + 
        \lt( B-A \rt) \sinh \frac{\dg t}{2}   \rt] \! , \label{twoex2}
\end{eqnarray}
but fitted to a single exponential 
{ 
\begin{eqnarray}
 F\lt[ f, t \rt] &=& 
        { \Gamma_f} \,   e^{-\Gamma_f \, t}. \label{singex} 
\end{eqnarray}
{ In \eq{twoex2} we have averaged over $f$ and its CP-conjugate
$\ov{f}$.} 
Of course the coefficients depend on the final state: 
$A=A(f)$, $B=B(f)$.}
A maximum likelihood fit { of \eq{singex} converges} to \cite{hm} 
\begin{eqnarray}
\Gamma_f &=& 
        \frac{A/\Gamma_L + B/\Gamma_H}{A/\Gamma_L^2 + B/\Gamma_H^2}     
 . \label{fitex}
\end{eqnarray}
We expand this to second order in \dg:
{
\begin{eqnarray}
     \Gamma_f &=& \Gamma \, + \,  
\frac{A-B}{A+B} \, \frac{\dg}{2} \, 
                - \frac{2 \, A B}{(A+B)^2}  \frac{(\dg)^2}{\Gamma} \,  
        + \, {\cal O} \lt( \frac{(\dg)^3}{\Gamma^2} \rt)
 . \label{fit}
\end{eqnarray}
}In flavour-specific decays we have $A=B$ (see \eq{guntf2}). We see
from \eq{fit} that here a single-exponential fit determines $\Gamma$
up to corrections of order $\dg^2/\Gamma^2$. 

Alternatively, one can use further theoretical input and exploit that
$\Gamma_{B_s}/\Gamma_{B_d}=1+{\cal O} (1\%) $ from heavy quark
symmetry \cite{bbsuv,bbd,kn}. This relation can therefore be used to
pinpoint $\Gamma$ in terms of the well-measured $B_d$ lifetime.  {
New physics in the standard penguin coefficients of the effective
$\Delta B=1$ hamiltonian only mildly affects
$\Gamma_{B_s}/\Gamma_{B_d}$ \cite{kn}. The full impact of new physics
on $\Gamma_{B_s}/\Gamma_{B_d}$, however, has not been studied yet.}

With \eq{guntf2} and \eq{twoex2} we can read off $A$ and $B$ for 
the CP-specific decays of Table~\ref{tab} 
{ 
and find $A(f_{\rm CP+})/B(f_{\rm CP+}) = (1+\cos \phi)/(1-\cos \phi)$ and 
$A(f_{\rm CP-})/B(f_{\rm CP-}) = (1-\cos \phi)/(1+\cos \phi)$ for CP-even and
CP-odd final states, respectively. Our key quantity for the discussion 
of CP-specific decays $\Bsun \to f_{\rm CP}$ is 
\begin{eqnarray}
\dg^\prime_{\rm CP} &\equiv& - \eta_f \adg \, \dg \, 
                \; = \; 
              \dg \, \cos \phi 
          \; = \; 
          \dg_{\rm CP} \, \cos^2 \phi  
        .\label{dgp}
\end{eqnarray}   
With this definition \eq{fit} reads for the decay rate $\Gamma_{\rm CP,
  \eta_f}$ 
measured in $\Bsun \to f_{\rm CP}$: 
\begin{eqnarray} 
     \Gamma_{\rm CP,\eta_f} &=& \Gamma \, + \,  
        \eta_f \, \frac{\dg^\prime_{\rm CP}}{2} 
        { \, - \, \sin^2 \phi \, \frac{(\dg)^2}{2 \Gamma} }
        \, + \, {\cal O} \lt( \frac{(\dg)^3}{\Gamma^2} \rt) . \no 
\end{eqnarray}
That is, to first order in \dg, comparing the $\Bsun$ lifetimes
measured in a flavour-specific and a CP-specific final state determines 
$\dg^\prime_{\rm CP}$.} 
Our result agrees with the one in \cite{g},
which has found \eq{dgp} by expanding the time evolution in
\eq{twoex2} and \eq{singex} for small $\dg \, t$.
Including terms of order $(\dg)^2$, lifetime measurements in 
a flavour-specific decay $\Bsun \to f_{\rm fs}$ determine \cite{hm}: 
\begin{eqnarray}
\Gamma_{\rm fs} &=& \Gamma \, - \, \frac{(\dg)^2}{2\Gamma} 
        \, + \, {\cal O} \lt( \frac{(\dg)^3}{\Gamma^2} \rt) . \no 
\end{eqnarray}
This implies $\Gamma_{\rm fs} < \Gamma$.  Despite the heavy quark symmetry
prediction $\Gamma_{B_s}/\Gamma_{B_d}\simeq 1 $, a large \dg\ leads to
an excess of the $B_s$ lifetime measured in $\Bsun \to f_{\rm fs}$ over
the $B_d$ lifetime \cite{hm}. From \eq{fit} one finds
\begin{eqnarray}
\Gamma_{\rm CP,\eta_f}  \, - \, \Gamma_{\rm fs}    & = &
        { \frac{\dg^\prime_{\rm CP}}{2}} \lt( \eta_f \, + \, 
                \frac{\dg^\prime_{\rm CP}}{\Gamma} \rt) \, + \, 
        {\cal O} \lt( \frac{(\dg)^3}{\Gamma^2}  \rt)  
        . \label{dgp2}
\end{eqnarray}
Hence for a CP-even (CP-odd) final state the quadratic corrections
enlarge (diminish) the difference between the two measured widths.  A
measurement of $\dg^\prime_{\rm CP}$ at Run-II of the Tevatron seems to be
feasible.  The lifetime measurement in the decay mode $\Bsun \to
J/\psi \phi$ has been studied in simulations \cite{sim,mm}.  This
decay mode requires an angular analysis to separate the CP-odd
(P-wave) from the CP-even (S-wave and D-wave) components. The angular
analysis is discussed in sect.~\ref{BsVV}.  With 2 fb$^{-1}$
integrated luminosity CDF expects 4000 reconstructed $\Bsun \to
J/\psi[\to \mu \ov{\mu}] \phi$ events and a measurement of
$\dg^\prime_{\rm CP}/\Gamma$ with an { absolute} error of 0.052. This {
  simulation} assumes that $\Gamma -(\dg)^2/({ 2}\Gamma)$ (see
\eq{fit}) will be measured from flavour-specific decays with an
accuracy of 1\% \cite{mm} { and uses the input $\dg^\prime_{\rm CP}/\Gamma
  =0.15$.}  When combining this with other modes in Table~\ref{tab} and
taking into account that an integrated luminosity of 10--20 fb$^{-1}$
is within reach of an extended (up to 2006) Run-II, the { study} of
$\dg^\prime_{\rm CP}$ at CDF looks very promising.  {The LHC
  experiments ATLAS, CMS and LHCb expect to measure
  $\dg^\prime_{\rm CP}/\Gamma$ with absolute errors between 0.012 and
  0.018 for $\dg^\prime_{\rm CP}/\Gamma =0.15$ \cite{cern}.  An upper bound on
  $\dg^\prime_{\rm CP}$} would be especially interesting. If the lattice
  calculations entering \eq{dgnum2} mature and the theoretical
  uncertainty decreases, an upper bound on $|\dg^\prime_{\rm CP}|$ may {
    show that} $\phi \neq 0,\pi$ through
\begin{eqnarray}
\frac{\dg^\prime_{\rm CP}}{\dg_{\rm CP}}  &=& \cos^2 \phi  .\label{cp2}
\end{eqnarray}  
Note that conversely the experimental establishment of a non-zero
$\dg^\prime_{\rm CP}$ immediately helps to constrain models of new
physics, because it excludes values of $\phi$ around $\pi/2$. This
feature  even holds true, if there is no theoretical progress
in \eq{dgnum2}. 

The described method to obtain $\dg_{\rm CP}^\prime$ can also be used, if
the sample contains { a known ratio of} CP-even and CP-odd
components.  This situation occurs e.g.\ in decays to $J/\psi \phi$,
if no angular analysis is performed or in final states, which are
neither flavour-specific nor CP eigenstates. We discuss this case
below in sect.~\ref{sec:both} with $\Bsun \to D_s^\pm D_s^{(*)}{}^\mp
$.  A measurement of the $B_s$ lifetime in $\Bsun \to J/\psi \phi$ has
been performed in \cite{cdf}, but the error is still too large to gain
information on $\dg^\prime_{\rm CP}$. { Note that the comparison of the 
lifetimes measured in CP-even and CP-odd final states determines 
$\dg^\prime_{\rm CP}$ up to corrections of order $(\dg/\Gamma)^3$.} 

\boldmath
\subsubsection{Determination of $|\dg|$ and $|\cos \phi|$}\label{sec:both}
\unboldmath
The theoretical uncertainty in \eq{dgnum2} dilutes the extraction 
of $|\cos \phi|$ from a measurement of $\dg_{\rm CP}^\prime$ alone. 
One can bypass the theory prediction in \eq{dgnum2} altogether by
measuring both  $\dg_{\rm CP}^\prime$ and $|\dg|$ and determine 
$|\cos \phi|$ through
\begin{eqnarray}
\frac{\dg^\prime_{\rm CP}}{\lt| \dg \rt|} &=& |\cos \phi | . \label{cp3}
\end{eqnarray}
To obtain additional information on \dg\ and $\phi$ from the time
evolution in \eq{guntf2} requires more statistics: the coefficient of
$t$ in \eq{guntf3}, $\dg\, \adg/2$, vanishes in flavour-specific
decays and is equal to $-\eta_f \dg_{\rm CP}^\prime/2 $ in the CP-specific
decays of Table~\ref{tab}. Therefore the data sample must be large
enough to be sensitive to the terms of order $(\dg \, t)^2$ in order
to get new information on \dg\ and $\phi$.  We now list { three} methods
to determine $|\dg|$ and $|\cos \phi|$ separately. The theoretical
uncertainty decreases and the required experimental statistics
increases from method 1 to method { 3}. Hence as the collected data
sample grows, one can work off our list downwards. The first method
exploits information from branching ratios and needs no information
from the quadratic $(\dg \,t)^2$ terms.

\paragraph{Method 1:} 
{ We assume that $\dg_{\rm CP}^\prime$ has been measured as described
in sect.~\ref{sec:dgcos}. The method presented now is a measurement of
$\dg_{\rm CP}$ using the information from branching ratios. With \eq{cp2}
one can then find $|\cos \phi|$ and subsequently $|\dg|$ from
\eq{cp3}.}  In the SV limit the branching ratio $\brunt{D_s^{(*)}{}^+
D_s^{(*)}{}^-}$ equals $\dg_{\rm CP}/(2 \Gamma)$ { up to corrections of
order $\dg/\Gamma$,} as discussed in sect.~\ref{sec:wd} \cite{ayopr}.
Corrections to the SV limit, however, can be sizeable. Yet we stress
that one can control the corrections to this limit experimentally, 
successively arriving at a result which does not rely on the validity of
the SV limit.  { For this it is of prime importance to determine the
CP-odd component of the final states $D_s^{\pm} D_s^{*\mp}$ and
$D_s^{*+} D_s^{*-}$. We now explain how the CP-odd and CP-even
component of any decay $\Bsun \to f$ corresponding to the quark level
transition $b \to c\ov{c} s$ can be obtained. This simply requires a
fit of the time evolution of the decay to a single exponential, as in
\eq{singex}.  { Define the contributions of the CP-odd and CP-even
eigenstate to $B_s \to f$:
\begin{eqnarray}
\Gamma (B_s^{\textrm{\scriptsize odd}} \to f) \; \equiv \; {\cal N}_f 
 \, |  \langle f
         \ket{B_s^{\textrm{\scriptsize odd}}} |^2,  
        && \qquad \qquad 
\Gamma (B_s^{\textrm{\scriptsize even}} \to f) \; \equiv \; {\cal N}_f 
        \, | \langle f
         \ket{B_s^{\textrm{\scriptsize even}}} |^2 . \label{defoe}
\end{eqnarray}
It is useful to define the CP-odd fraction $x_f$ by
\begin{eqnarray}
\frac{  \Gamma (B_s^{\textrm{\scriptsize odd}}  
                \to f) }{
        \Gamma ( B_s^{\textrm{\scriptsize even}} 
                \to f) }
& = & \frac{  \lt|  \langle f
         \ket{B_s^{\textrm{\scriptsize odd}}} \rt|^2 }{
        \lt| \langle f
         \ket{B_s^{\textrm{\scriptsize even}}} \rt|^2 } \; =\; 
\frac{  \lt|  \langle \ov{f}
         \ket{B_s^{\textrm{\scriptsize odd}}} \rt|^2 }{
        \lt| \langle \ov{f}
         \ket{B_s^{\textrm{\scriptsize even}}} \rt|^2 } \; =\;
  \frac{x_f}{1-x_f}  . \label{x1x}
\end{eqnarray}
The time evolution $(\guntf +\guntfb)/2$ of the CP-averaged untagged
decay $\Bsun \to f,\ov{f}$ is governed by a two-exponential formula:
\begin{eqnarray}
\frac{\guntf +\guntfb }{2} &=& 
        A(f) \, e^{-\Gamma_L t} + B(f) \, e^{-\Gamma_H t} .\label{twoex3} 
\end{eqnarray}
With \eq{rot} and \eq{twoex} one finds  
\begin{eqnarray}
A(f) & = & 
 \frac{{\cal N}_f}{2} \,  | \langle  f \ket{B_L} |^2 \, +  
 \frac{{\cal N}_f}{2} \,        | \langle  \ov{f} \ket{B_L} |^2 \nn
  & = & \frac{1+ \cos \phi}{2} \, 
        \Gamma ( B_s^{\textrm{\scriptsize even}} \to f) \, + \,
        \frac{1- \cos \phi}{2} \, 
        \Gamma ( B_s^{\textrm{\scriptsize odd}} \to f) \nn
B(f) & = & \frac{{\cal N}_f}{2} \,
         | \langle f  \ket{B_H } |^2 \, + \, 
        \frac{{\cal N}_f}{2} \, | \langle \ov{f}  \ket{B_H } |^2 \nn
  & = & \frac{ 1- \cos \phi }{2} \, 
        \Gamma ( B_s^{\textrm{\scriptsize even}} \to f) \, + \, 
          \frac{ 1+ \cos \phi }{2} \, 
        \Gamma ( B_s^{\textrm{\scriptsize odd}} \to f) 
        .\label{abg}
\end{eqnarray}
With \eq{x1x} we arrive at
\begin{eqnarray}
\frac{A(f)}{B(f)} & = & 
 \frac{ (1+ \cos \phi) 
        \Gamma ( B_s^{\textrm{\scriptsize even}} \to f) + 
        (1- \cos \phi) 
        \Gamma ( B_s^{\textrm{\scriptsize odd}} \to f)}{
        (1- \cos \phi) 
        \Gamma ( B_s^{\textrm{\scriptsize even}} \to f) + 
        (1+ \cos \phi) 
        \Gamma ( B_s^{\textrm{\scriptsize odd}} \to f) 
        } 
        \; = \; 
        \frac{1+(1- 2 x_f) \cos \phi}{1-(1- 2 x_f) \cos \phi}
.\label{ab}
\end{eqnarray}
In \eq{abg} and \eq{ab} it is crucial that we average the decay rates
for $\Bsun \to f$ and the CP-conjugate process $\Bsun \to
\ov{f}$. This eliminates the interference term
$\bra{B_s^{\textrm{\scriptsize odd}}} f\rangle \bra{f}
B_s^{\textrm{\scriptsize even}} \rangle $, so that $A(f)/B(f)$ only depends
on $x_f$.  The single exponential fit with \eq{singex} determines
$\Gamma_f$. Equations \eq{fit} and \eq{ab} combine to give
\begin{eqnarray}
{ 2\, ( \Gamma_f - \Gamma )} &=& 
        (1- 2 x_f ) \, \dg \, \cos \phi 
  \; = \; (1- 2 x_f ) \, \dg_{\rm CP} \, \cos^2 \phi
  \; = \; (1- 2 x_f ) \, \dg_{\rm CP}^\prime 
 \label{dgmx} 
\end{eqnarray}
up to corrections of order $(\dg)^2/\Gamma$. In order to determine
$x_f$ from \eq{dgmx} we need $\dg_{\rm CP}^\prime$ from the lifetime
measurement in a CP-specific final state like $D_s^+ D_s^-$ or from
the angular separation of the CP components in $\Bsun \to \psi
\phi$. The corrections of order $(\dg)^2/\Gamma$ to
\eq{dgmx} can be read off from \eq{fit} with \eq{ab} as well.  {
Expressing the result in terms of $\Gamma_f$ and the rate $\Gamma_{\rm
fs}$ measured in flavour-specific decays, we find
\begin{eqnarray}
1- 2 x_f &=& 2\, 
        \frac{\Gamma_f-\Gamma_{\rm fs}}{\dg_{\rm CP}^\prime} 
        \, \lt[ 1 \, -  \,  
        2\, \frac{\Gamma_f-\Gamma_{\rm fs}}{\Gamma} \rt] \, 
        + {\cal O} \lt( \frac{(\dg)^2}{\Gamma^2} \rt)
        . \label{xfres} 
\end{eqnarray} 
}In order to solve for $\Gamma (B_s^{\textrm{\scriptsize even}} \to f)
$ and $\Gamma (B_s^{\textrm{\scriptsize odd}} \to f) $ we also need
the branching ratio $\brunt{f}+\brunt{\ov{f}}$. Recalling 
\eq{untnorm} one finds from \eq{twoex3} and \eq{abg}:
\begin{eqnarray}
\brunt{f}+\brunt{\ov{f}} 
&=& \phantom{ + }
\Gamma (B_s^{\textrm{\scriptsize even}}  \to f)
  \lt[    \frac{1+ \cos \phi }{2 \Gamma_L}  + 
         \frac{1- \cos \phi }{2 \Gamma_H} \rt]  \nn
&&  + \, 
\Gamma (B_s^{\textrm{\scriptsize odd}}  \to f)
  \lt[    \frac{1- \cos \phi }{2 \Gamma_L}  + 
         \frac{1+ \cos \phi }{2 \Gamma_H} \rt] 
. \label{boe}
\end{eqnarray}
By combining \eq{x1x} and \eq{boe} we can solve for the two CP components:
\begin{eqnarray}
\Gamma (B_s^{\textrm{\scriptsize even}} \to f)
&=& 
        \lt[ \Gamma^2- \lt( \dg/2 \rt)^2 \rt] \, 
        \lt( \brunt{ f} + \brunt{\ov{f}} \rt)   \,      
   \frac{1- x_f}{ 2 \Gamma \, - \, \Gamma_f } \nn
& = &   (1- x_f) \, \lt( \brunt{ f} + \brunt{\ov{f}} \rt)       \,
        \Gamma 
        +\, {\cal O} \lt( \dg  \rt) \nn  
\Gamma (B_s^{\textrm{\scriptsize odd}}  \to f)
&=&  
        \lt[ \Gamma^2- \lt( \dg/2 \rt)^2 \rt] \, 
        \lt( \brunt{ f} + \brunt{\ov{f}} \rt)   \,
   \frac{x_f}{ 2 \Gamma \, - \, \Gamma_f } \nn
& = &   x_f \,  \lt( \brunt{ f} + \brunt{\ov{f}} \rt)   \,
         \Gamma \,  
        +\, {\cal O} \lt( \dg \rt) \no .
\end{eqnarray}
From \eq{dgcp2} we now find the desired quantity by summing over all 
final states $f$:
\begin{eqnarray} 
\dg_{\rm CP} \; = \; 
\Gamma \lt( B_s^{\textrm{\scriptsize even}} \rt) - 
        \Gamma \lt( B_s^{\textrm{\scriptsize odd}} \rt) 
&=&
        2 \,   \lt[ \Gamma^2- \lt( \dg/2 \rt)^2 \rt]
\sum_{f \in X_{c\ov{c}}}  \brunt{ f}    \,
   \frac{1- 2\, x_f}{ 2 \Gamma \, - \, \Gamma_f} 
  \label{dgcpres} \\
&=& 2\,  \Gamma \sum_{f \in X_{c\ov{c}}}  \brunt{ f} 
        \, 
        (1- 2\, x_f)\, \lt[ 1 \, + \, {\cal O} \lt( \frac{\dg}{\Gamma}
        \rt) \rt].
\label{dgcpres2} 
\end{eqnarray}
{ It is easy to find $\dg_{\rm CP}$: first determine $1-2x_f$ from
\eq{xfres} for each studied decay mode, then insert the result into
\eq{dgcpres}. The small quadratic term $( \dg/2 )^2=\dg_{\rm CP}
\dg_{\rm CP}^\prime/4$ is negligible.}  This procedure can be performed
for $\brunt{D_s^{\pm} D_s^{*}{}^{\mp}}$ and $\brunt{D_s^{*}{}^+
D_s^{*}{}^-}$ to determine the corrections to the SV limit. In
principle the CP-odd P-wave component of $\brunt{D_s^{*}{}^+
D_s^{*}{}^-}$ (which vanishes in the SV limit) could also be obtained
by an angular analysis, but this is difficult in first-generation
experiments at hadron colliders, because the photon from
$D_s^* \to D_s \gamma$ cannot be detected.  We emphasize that it is
not necessary to separate the $D_s^{(*)}{}^+D_s^{(*)}{}^-$ final
states; our method can also be applied to the semi-inclusive
$D_s^{(*)}{}^\pm D_s^{(*)}{}^\mp $ sample, using 
$\dg_{\rm CP}^\prime$ obtained from an angular separation of the CP
components in $\Bsun \to \psi \phi$.  Further one can successively
include those double-charm final states which vanish in the SV limit
into \eq{dgcpres}.  If we were able to reconstruct all $b \to c \ov{c}
s$ final states, we could determine $\dg_{\rm CP}$ without invoking the SV
limit.  In practice a portion of these final states will be missed,
but the induced error can be estimated from the corrections to the SV
limit in the measured decay modes. By comparing $\dg_{\rm CP}$ and
$\dg_{\rm CP}^\prime$ one finds $|\cos \phi|$ from \eq{cp2}.  } 
{ The irreducible theoretical error of method 1 stems from the
omission of CKM-suppressed decays and is of order $2 |V_{ub}
V_{us}/(V_{cb} V_{cs})| \sim 5\%$.} 

Method 1 is experimentally simple: at the first stage (relying on the
SV limit) it amounts to counting the $\Bsun$ decays into $
D_s^{(*)}{}^+ D_s^{(*)}{}^-$. A first simulation indicates that CDF
will be able to separate the $B_s$ decay modes into $D_s^+ D_s^-$, {
  $D_s^{*\pm} D_s^{\mp}$} and $D_s^{*+} D_s^{*-}$ \cite{cp}.  The
corrections to the SV limit are obtained by one-parameter fits to the
time evolution of the collected double-charm data samples.  This
sample may include final states from decay modes which vanish in the
SV limit, such as multiparticle final states.  No sensitivity to $(\dg
\, t)^2$ is needed.  A further advantage is that $\dg_{\rm CP}$ is not
diminished by the presence of new physics.

% \paragraph{Method 2:} 
% In \cite{hm} the lifetime $1/{ \Gamma_f}$ measured by
% fitting \eq{singex} to the decay time evolution in flavour-specific
% decays has been studied. The result has been compared to the measured
% $B_d$ lifetime $\tau_{B_d}$ in order to constrain $|\dg|$ from the
% second term in \eq{fit} (in which $A=B$).  One can use this method to
% obtain $|\dg|$ and then $|\cos \phi|$ from \eq{cp3}.  The theoretical
% input is $\Gamma_{B_s}/\Gamma_{B_d}=1+{\cal O} (1\%) $ from heavy
% quark symmetry \cite{bbsuv,bbd,kn}. This method only involves a
% one-parameter fit, but is quadratic in $\dg$.  { Three
% different decays must be studied (to measure $\tau_{B_d}$,
% $\tau_{B_s}$ and $\dg_{\rm CP}^\prime$) in order to find $|\cos \phi|$.}

\paragraph{Method 2:}
In the Standard Model the decay into a CP eigenstate $f_{\rm CP}$ is
governed by a single exponential. If a second exponential is found in
the time evolution of a CKM-favoured decay $\Bsun \to f_{\rm CP}$, this
will be clear evidence of new physics \cite{dun}. To this end we must
resolve the time evolution in \eq{guntf2} up to order $(\dg \,
t)^2$. At first glance this seems to require a three-parameter fit to
the data, because \guntf\ in \eq{guntf2} depends on $\Gamma$, $\dg$
and (through $\adg$, see \eq{acp2}) on $\phi$. It is possible,
however, to choose these parameters in such a way that one of them
enters \gunt{f_{\rm CP}}\ at order $(\dg )^3$, with negligible impact.
The fit parameters are $\Gamma^\prime$ and $Y$.  They are chosen such
that
\begin{eqnarray} 
\!\!\!\!
\guntfcpp &=&  2\, \brunt{f_{\rm CP+} } \,\,
        \Gamma^\prime e^{-\Gamma^\prime t} \, 
        \lt[ 1 + Y \, \Gamma^\prime \, t \, 
        \lt( -1 
                +\frac{\Gamma^\prime t}{2} \rt)
                + {\cal O} \lt( (\dg )^3 \rt)
         \rt] 
        . \label{t2}
\end{eqnarray}
Here we have considered a CP-even final state, for which a lot more
data are expected than for CP-odd states.  With \eq{t2} we have
generalized the lifetime fit method described in sect.~\ref{sec:dgcos}
to the order $(\dg\, t)^2$.  A non-zero $Y$ signals the presence of
new physics. 
The fitted rate $\Gamma^\prime$ and $Y$ are related to
$\Gamma$, $\dg$ and $\phi$ by
\begin{eqnarray} 
Y \; = \; \frac{(\dg)^2}{4 \Gamma^{\prime 2}} \sin^2 \phi 
        ,  \qquad \qquad \qquad
\Gamma^\prime &=& \Gamma (1-Y) + \frac{\cos \phi}{2} \dg 
      \label{defgpy} .
\end{eqnarray} 
Note that for $|\cos \phi|=1$ the rate $\Gamma^\prime$ equals the 
rate of the shorter-lived mass eigenstate and the expansion in \eq{t2} 
becomes the exact single-exponential formula.  
{ After determining $\Gamma^\prime$ and $Y$ we can solve \eq{defgpy}
for $\Gamma$, $\dg$ and $\phi$. To this end we need the width
$\Gamma_{\rm fs}$ measured in flavour-specific decays. We find
\begin{eqnarray}
|\dg| \;=\; 2 \sqrt{(\Gamma^\prime-\Gamma_{\rm fs})^2 +\Gamma_{\rm fs}^2 Y } 
        \lt[ 1+ {\cal O} \lt( \frac{\dg}{\Gamma}  \rt) \rt], 
&& \quad 
\Gamma\; = \; \Gamma_{\rm fs} + \frac{(\dg)^2}{2 \Gamma} + 
                {\cal O} \lt( \lt(\frac{\dg}{\Gamma}\rt)^3  \rt) \nn 
\dg_{\rm CP}^\prime \; =\; 2 \lt[ \Gamma^\prime - \Gamma \lt( 1- Y \rt) \rt] 
        \lt[ 1 + {\cal O} \lt( \lt(\frac{\dg}{\Gamma}\rt)^2  \rt)  \rt],  
&& \quad 
|\sin \phi| \; = \; \frac{2\Gamma \sqrt{Y}}{|\dg|} \, 
        \lt[ 1+ {\cal O} \lt( \frac{\dg}{\Gamma}  \rt) \rt] \!. 
        \label{gpyres}
\end{eqnarray} 
The { quantity} $\dg_{\rm CP}^\prime$, which we could already 
determine from single-exponential fits, { is} now found beyond the
leading order in $\dg/\Gamma$. By contrast, \dg\ and $|\sin \phi|$ in 
\eq{gpyres} are only determined to the first non-vanishing order in 
$\dg/\Gamma$.} 

In conclusion method 2 involves a two-parameter fit and needs
sensitivity to the quadratic term in the time evolution. The presence
of new physics can be invoked from $Y\neq 0$ and does not require to
combine lifetime measurements in different decay modes. 

\paragraph{Method 3:}
Originally the following method has been proposed to determine $|\dg|$
\cite{dun,g}: The time evolution of a $\Bsun$ decay into a
flavour-specific final state is fitted to two exponentials. This
amounts to resolving the deviation of $\cosh (\dg\, t/2)$ from 1 in
\eq{guntf2} in a two-parameter fit for $\Gamma$ and $|\dg|$. If one
adopts the same parametrization as in \eq{t2}, $\Gamma^\prime$ and
$Y$ are obtained from \eq{defgpy} by replacing $\phi$ with $\pi/2$.
The best suited flavour-specific decay modes at hadron colliders are
$\Bsun \to D_s^{(*)\pm} \pi^\mp $, $\Bsun \to D_s^{(*)\pm} \pi^\mp
\pi^+ \pi^- $ and $\Bsun \to D_s^{(*)\pm} X \ell^{\mp} \nu$.
Depending on the event rate in these modes, method 3 could be superior
to method 2 in terms of statistics. On the other hand, to find the
``smoking gun'' of new physics, the $|\dg|$ obtained  must be
compared to $\dg_{\rm CP}^\prime$ from CP-specific decays to prove $|\cos
\phi|\neq 1$ through \eq{cp3}. Since the two measurements are
differently affected by systematic errors, this can be a difficult
task. First upper bounds on $|\dg|$ using method 3 have been obtained
in \cite{semi}.

The L3 collaboration has determined an upper bound $|\dg|/\Gamma\leq
0.67$ by fitting the time evolution of fully inclusive decays to two
exponentials \cite{l3}. This method is quadratic in $\dg$ as well. The
corresponding formula for the time evolution can be simply obtained
from \eq{twoex2} with $A=\Gamma_L$ and $B=\Gamma_H$.
 
\subsection{ CP Violation in Mixing and 
  Untagged Oscillations}\label{sec:cpmix} 
In the preceding sections we have set the small parameter $a$ in
\eq{defa} to zero. { CP violation in mixing} vanishes in this limit.
The corresponding ``wrong-sign'' CP asymmetry is measured in
flavour-specific decays and equals
\begin{eqnarray}
a_{\rm fs} &=& \frac{\gbtf - \gtfb}{\gbtf + \gtfb} \; = \; a
        \quad \qquad \mbox{for} \quad \ov{A}_f  =  0 
        \quad \mbox{and} \quad |A_f| = |\ov{A}_{\ov{f}}|
 . \label{defafs}
\end{eqnarray}
{ A special case of $a_{\rm fs}$ is the semileptonic asymmetry, where
$f=X \ell^+ \nu$.}
{ A determination of $a$ gives additional information on the three
physical quantities $|M_{12}|$, $|\Gamma_{12}|$ and $\phi$
characterizing \bbms. Measuring $\dm$, $\dg_{\rm CP}$, $\dg_{\rm CP}^\prime$
and $a$ overconstrains these quantities. 

The ``right-sign'' asymmetry vanishes:}
\begin{eqnarray}
    \gtf - \gbtfb \; = \; 0
        \quad \qquad \mbox{for} \quad \ov{A}_f  =  0 
        \quad \mbox{and} \quad |A_f| = |\ov{A}_{\ov{f}}|
 . \label{unm}
\end{eqnarray}
This implies that one can measure $a_{\rm fs}$ from \emph{untagged}
decays. { This observation was already made in \cite{y}}. It is
easily verified from the sum of \eq{gtfres} and \eq{gbtfres} that to
order $a$ the time evolution of untagged decays exhibits oscillations
governed by \dm.  Since $a$ is small, one must be concerned to which
accuracy $|A_f| = |\ov{A}_{\ov{f}}|$ holds in flavour-specific decays
in the presence of new physics. For example in left--right-symmetric
extensions of the Standard Model, small CP-violating corrections to the
decay amplitude could eventually spoil this relation at the few
per mille level. Further, a small production asymmetry $\epsilon =
N_{\ov{B}}/N_B -1$ also leads to oscillations in the untagged sample.
To first order in the small parameters $a$, $\epsilon$ and
$|A_f|/|\ov{A}_{\ov{f}}|-1$ one finds
\begin{eqnarray}
a_{\rm fs}^{unt} &=& 
   \frac{\guntf - \guntfb}{\guntf + \guntfb} 
        \nn 
& = &
  \frac{|A_f|^2-|\ov{A}_{\ov{f}}|^2}{|A_f|^2+|\ov{A}_{\ov{f}}|^2} 
        \, + \, 
        \frac{a}{2} - \frac{a+\epsilon}{2} \,  
        \frac{\cos (\dm\, t)}{\cosh (\dg t/2) }
        \qquad \mbox{for} \quad \ov{A}_f  =  0 
        \quad \mbox{and} \quad |A_f| \approx |\ov{A}_{\ov{f}}| 
        . \,  \label{fsun} 
\end{eqnarray}
For $|A_f| = |\ov{A}_{\ov{f}}|$ and $\epsilon=0$ one recovers the
formula derived in \cite{y}.}  Note that the production asymmetry
between $B_s$ and $\ov{B}_s$ cannot completely fake the effect of a
non-zero $a$ in \eq{fsun}: while both $a\neq 0$ and $\epsilon\neq 0$
lead to oscillations, the offset from the constant term indicates new
CP-violating physics either in \bbms\ (through $a\neq 0$) or in the
studied decay amplitude (through $|A_f|\neq |\ov{A}_{\ov{f}}|$). The
latter effect, which is theoretically { much} less likely, can be
tested in $B^\pm$ decays and can therefore be disentangled from $a\neq 0$.

The ratio $\dg_{\rm CP}/\Gamma \leq 0.22$ from \eq{dgnum2} and the current
experimental limit $\dm \geq { 14.9}\,$ps${}^{-1}$ \cite{os} imply that
$|a| \leq 0.01$.  CDF expects sufficiently many reconstructed $\Bsun
\to D_s^{(*)\pm} \pi^{\mp}$ and $\Bsun \to D_s^{(*)\pm} \pi^{\mp}
\pi^+ \pi^-$ events at Run-II after collecting 2 fb$^{-1}$ of integrated
luminosity to achieve a statistical error at the few permille level.
{ From \eq{defa} and \eq{dmdg} we can relate $a$ to $|\dg|$, \dm\
and $\phi$:}
\begin{eqnarray}
a  &=& \frac{|\dg|}{\dm} \, \frac{\sin \phi}{|\cos \phi|} . \no
\end{eqnarray}
Note, however, that the measurement of the sign of $a$ determines the
sign of $\sin \phi$. This reduces the four-fold ambiguity in $\phi$
from the measurement of $|\cos \phi| $ to a two-fold one. It is
interesting that, at order $a$, without tagging one can { in
principle} gain information which otherwise requires tagged
studies. Of course $\sin \phi$ can be measured more directly from
tagged decays, as discussed in the forthcoming section
\ref{sec:tagged}.

\section{Tagged Decays}\label{sec:tagged}
\boldmath
\subsection{The CP-Violating Observables of $B_s\to D_s^+D^-_s$ and
$J/\psi\, \eta^{(\prime)}$}\label{BsPP}
\unboldmath
For a $B_s$ decay into a CP eigenstate $f$ 
the $B_s$--$\overline{B_s}$ oscillations lead to the following 
time-dependent CP asymmetry:
\begin{eqnarray}
a_{\rm CP}(t) & \equiv &
 \frac{ \Gamma(\overline{B_s}(t)\to f) - \Gamma(B_s(t)\to f)
      }{\Gamma(B_s(t)\to f)+ \Gamma(\overline{B_s}(t)\to f) }
\; = \;  
        -\; \frac{ \adi  \cos(\dm \, t) + \ami  \sin(\dm \, t)}{
          \cosh \lt( \dg \, t/2 \rt) + \adg \sinh \lt( \dg \, t/2 \rt)     
             } 
        .\label{ee6}
\end{eqnarray}
Here the mass and width difference $\dm$ and $\dg$ can be found in
\eq{defdm} and $\adi$, $\ami$ and $\adg$ have been defined in
\eq{defacp}.  We have set the small parameter $a$ in \eq{defa} to zero
and will continue to do so.  The final states $B_s\to
D_s^+D^-_s$, { $\psi\, \eta^{(\prime)}$, $\psi f_0$} or $\chi_{c0} \phi$ in
Table~\ref{tab} are CP eigenstates.  Their CP eigenvalue $\eta_f$ reads
$\eta_{D_s^+D^-_s}=\eta_{\psi \eta^\prime}=\eta_{\psi\eta}=+1$ and
${ \eta_{\psi f_0}=} \eta_{\chi_{c0} \phi}= -1$. With \eq{acp2} we then 
find from \eq{ee6}:
\begin{eqnarray}
a_{\rm CP}(t) & = &
        -\; \frac{ \eta_f \sin \phi \,  \sin(\dm \, t)}{
          \cosh \lt( \dg \, t/2 \rt) - 
              \eta_f |\cos \phi| \sinh \lt( |\dg| \, t/2 \rt)     
             } 
        .\label{aph}
\end{eqnarray}
Since $\dg$ and $\cos \phi$ have the same sign (see \eq{dgcph}) we
could replace these quantities by their absolute values in the
denominator of \eq{aph}. This displays that the ambiguity in the sign
of $\cos \phi$ cannot be removed by measuring $a_{\rm CP}$.  Its
measurement determines $\sin \phi$ and leaves us with a two-fold
ambiguity in $\phi$. Then we still do not know whether the heavier or
lighter mass eigenstate is shorter-lived.  The resolution of this
ambiguity will be discussed in Section~\ref{sec:ambig}.
\boldmath
\subsection{The CP-violating Observables of $B_s\to J/\psi\, \phi$
and $D_s^{\ast+}D_s^{\ast-}$}\label{BsVV}
\unboldmath
The situation in the decay $B_s\to J/\psi\, \phi$, which is very
promising for $B$-physics experiments at hadron machines because of
its nice experimental signature, is a bit more involved than in the
case of the pseudoscalar--pseudoscalar modes $B_s\to D_s^+D^-_s$ and
$J/\psi\, \eta^{(\prime)}$, since the final state is an admixture of
different CP eigenstates. In order to disentangle them, we have to
make use of the angular distribution of the decay products of the
decay chain $B_s\to J/\psi[\to l^+l^-]\, \phi[\to K^+K^-]$, which can
be found in \cite{ddlr,ddf1}. In that paper, also appropriate
weighting functions are given to extract the observables of the
angular distribution in an efficient way from the experimental
data. For an initially, i.e.\ at time $t=0$, present $B_s$-meson, the
time-dependent angular distribution can be written generically as
\begin{equation}\label{ang}
f(\Theta,\Phi,\Psi;t)=\sum_k{\cal O}^{(k)}(t)\,
g^{(k)}(\Theta,\Phi,\Psi),
\end{equation}
where we have denoted the angles describing the kinematics of the
decay products of $J/\psi\to l^+l^-$ and $\phi\to K^+K^-$ by $\Theta$,
$\Phi$ and $\Psi$. The observables ${\cal O}^{(k)}(t)$ describing the
time evolution of the angular distribution (\ref{ang}) can be
expressed in terms of real or imaginary parts of certain bilinear
combinations of decay amplitudes. In the case of decays into two
vector mesons, such as $B_s\to J/\psi\, \phi$, it is convenient to
introduce linear polarization amplitudes $A_0(t)$, $A_\parallel(t)$
and $A_\perp(t)$ \cite{pol}. Whereas $A_\perp(t)$ describes a CP-odd
final-state configuration, both $A_0(t)$ and $A_\parallel(t)$
correspond to CP-even final-state configurations.  The observables
${\cal O}^{(k)}(t)$ of the corresponding angular distribution are
given by
\begin{equation}\label{obs1}
\left|A_f(t)\right|^2\quad\mbox{with}\quad f\in\{0,\parallel,\perp\},
\end{equation}
as well as by the interference terms
\begin{equation}\label{obs2}
\Re\{A_0^\ast(t)A_\parallel(t)\}\quad\mbox{and}\quad
\Im\{A_f^\ast(t)A_\perp(t)\} \quad\mbox{with}\quad f\in\{0,\parallel\}.
\end{equation}
For our consideration, the time evolution of these observables plays
a crucial role. In the case of the observables (\ref{obs1}), which 
correspond to ``ordinary'' decay rates, we obtain
\begin{eqnarray}
|A_0(t)|^2 &=&  
   |A_0(0)|^2 e^{-\Gamma t} \lt[ 
        \cosh \frac{\dg \, t}{2} \, - \, 
        |\cos \phi| \sinh \frac{|\dg| \, t}{2} \, + \, 
        \sin \phi \, \sin (\dm\, t) \rt] 
        \label{EQB}\\
|A_{\|}(t)|^2 &=& 
   |A_{\|}(0)|^2 e^{-\Gamma t} \lt[ 
        \cosh \frac{\dg \, t}{2} \, - \, 
        |\cos \phi| \sinh \frac{|\dg| \, t}{2} \, + \, 
        \sin \phi \, \sin (\dm\, t) \rt] 
        \\ 
|A_{\perp}(t)|^2 &=&
   |A_{\perp}(0)|^2 e^{-\Gamma t} \lt[ 
        \cosh \frac{\dg \, t}{2} \, + \, 
        |\cos \phi| \sinh \frac{|\dg| \, t}{2} \, - \, 
        \sin \phi \, \sin (\dm\, t) \rt] 
        , \label{EQP}
\end{eqnarray}
whereas we have in the case of the interference terms (\ref{obs2}):
\begin{eqnarray}
\!\!\!\!\!\!\! \Re\{A_0^*(t) A_{\|}(t)\} &=& 
   |A_0(0)| \, |A_{\|}(0)| \, \cos ( \delta_2 - \delta_1 ) \,
   e^{-\Gamma t} \nn 
  && \!\!\! \times \lt[ 
        \cosh \frac{\dg \, t}{2} \, - \, 
        |\cos \phi| \sinh \frac{|\dg| \, t}{2} \, + \, 
        \sin \phi \, \sin (\dm\, t) \rt]  \label{Re-expr}\\
\!\!\!\!\!\!\! \Im\{ A_{\|}^* (t) A_{\perp}(t)\} &=& 
   |A_{\|}(0)| \, |A_{\perp}(0)| \, e^{-\Gamma t} \nn
  && \!\!\! \times \lt[  
        \sin\delta_1 \, \cos(\dm\,  t) \, - \, 
        \cos\delta_1 \, \cos\phi \, \sin(\dm\,  t) \, - \, 
        \cos\delta_1 \, \sin\phi \, \sinh \frac{\dg \, t}{2} 
           \rt] \\
\!\!\!\!\!\!\! \Im\{A_{0}^*(t)A_{\perp}(t)\} &=& 
   |A_{0}(0)| \, |A_{\perp}(0)| \, e^{-\Gamma t} \nn 
  && \!\!\! \times \lt[  
        \sin\delta_2 \, \cos(\dm\,  t) \, - \, 
        \cos\delta_2 \, \cos\phi \, \sin(\dm\,  t) \, - \, 
        \cos\delta_2 \, \sin\phi \, \sinh \frac{\dg \, t}{2} 
           \rt] \! . \label{EQE}
\end{eqnarray}
In (\ref{Re-expr})--(\ref{EQE}), $\delta_1$ and $\delta_2$
denote CP-conserving strong phases, which are defined as follows 
\cite{ddlr,ddf1}:
\begin{equation}
\delta_1\equiv\mbox{arg}\Bigl\{A_{\|}(0)^\ast A_{\perp}(0)\Bigr\},\quad
\delta_2\equiv\mbox{arg}\Bigl\{A_{0}(0)^\ast A_{\perp}(0)\Bigr\}.
\end{equation}
The time evolutions (\ref{EQB})--(\ref{EQE}) generalize those given in
\cite{ddlr,ddf1} to the case of a sizeable \bbms\ phase $\phi$ to cover
the pursued case of new physics.   A further generalization taking
into account also the small penguin contributions can be found in
\cite{RF-ang}.  It should be emphasized that new physics manifests
itself {\it only} in the observables ${\cal O}^{(k)}(t)$, while
the $g^{(k)}(\Theta,\Phi,\Psi)$'s are not affected.

We may use the same angles $\Theta$, $\Phi$ and $\Psi$ to describe the
kinematics of the decay products of the CP-conjugate transition 
$\overline{B_s}\to J/\psi\, \phi$. Consequently, we have 
\begin{equation}\label{ang-CP}
\overline{f}(\Theta,\Phi,\Psi;t)=\sum_k\overline{{\cal O}}^{(k)}(t)\,
g^{(k)}(\Theta,\Phi,\Psi).
\end{equation}
Within this formalism, CP transformations relating 
$B_s\to\! [J/\psi\, \phi]_f$ to $\overline{B_s}\to\! [J/\psi\, \phi]_f$ 
\mbox{ ($f\! \in\! \{0,\parallel,\perp\}$)} are taken into account in the 
expressions 
for the ${\cal O}^{(k)}(t)$ and $\overline{{\cal O}}^{(k)}(t)$, 
and do not affect the form of the $g^{(k)}(\Theta,\Phi,\Psi)$. Therefore 
the same functions $g^{(k)}(\Theta,\Phi,\Psi)$ are present in (\ref{ang}) 
and (\ref{ang-CP}) (see also \cite{FD,FD1}). The CP-conjugate observables 
$\overline{{\cal O}}^{(k)}(t)$ take the following form:
\begin{eqnarray}
|\overline{A}_0(t)|^2 &=& 
   |A_0(0)|^2 e^{-\Gamma t} \lt[ 
        \cosh \frac{\dg \, t}{2} \, - \, 
        |\cos \phi| \sinh \frac{|\dg| \, t}{2} \, - \, 
        \sin \phi \, \sin (\dm\, t) \rt] \label{EQBb} \\
|\overline{A}_{\|}(t)|^2 &=& 
   |A_{\|}(0)|^2 e^{-\Gamma t} \lt[ 
        \cosh \frac{\dg \, t}{2} \, - \, 
        |\cos \phi| \sinh \frac{|\dg| \, t}{2} \, - \, 
        \sin \phi \, \sin (\dm\, t) \rt] 
        \\ 
|\overline{A}_{\perp}(t)|^2 &=& 
   |A_{\perp}(0)|^2 e^{-\Gamma t} \lt[ 
        \cosh \frac{\dg \, t}{2} \, + \, 
        |\cos \phi| \sinh \frac{|\dg| \, t}{2} \, + \, 
        \sin \phi \, \sin (\dm\, t) \rt] \label{EQPb}
\end{eqnarray}
\begin{eqnarray}
\!\!\!\!\!\!\! \Re\{\overline{A}_0^*(t) \overline{A}_{\|}(t)\} &=&
   |A_0(0)| \, |A_{\|}(0)| \, \cos ( \delta_2 - \delta_1 ) \,
   e^{-\Gamma t} \nn 
  && \! \times \lt[ 
        \cosh \frac{\dg \, t}{2} \, - \, 
        |\cos \phi| \sinh \frac{|\dg| \, t}{2} \, - \, 
        \sin \phi \, \sin (\dm\, t) \rt]  
\label{Re-exprb} \\
\!\!\!\!\!\!\! \Im\{\overline{A}_{\|}^*(t)\overline{A}_{\perp}(t)\} &=&
   |A_{\|}(0)| \, |A_{\perp}(0)| \, e^{-\Gamma t} \nn
  && \!\!\!\!\!\!\!\! \times \lt[  
        -\, \sin\delta_1 \, \cos(\dm\,  t) \, + \, 
        \cos\delta_1 \, \cos\phi \, \sin(\dm\,  t) \, - \, 
        \cos\delta_1 \, \sin\phi \, \sinh \frac{\dg \, t}{2} 
           \rt] \\
\!\!\!\!\!\!\! \Im\{\overline{A}_{0}^*(t)\overline{A}_{\perp}(t)\}&=&
   |A_{0}(0)| \, |A_{\perp}(0)| \, e^{-\Gamma t} \nn 
  && \!\!\!\!\!\!\!\! \times \lt[  
        -\, \sin\delta_2 \, \cos(\dm\,  t) \, + \, 
        \cos\delta_2 \, \cos\phi \, \sin(\dm\,  t) \, - \, 
        \cos\delta_2 \, \sin\phi \, \sinh \frac{\dg \, t}{2} 
           \rt] \!\! . \label{EQEb}
\end{eqnarray}
Note that one can determine $\sin \delta_{1,2}$, $ \cos
(\delta_1-\delta_2)$, $\sin \phi$, $ \cos \delta_i \cos \phi$, \dm\
and $|\dg|$ from (\ref{EQB})-(\ref{EQEb}). Using $\cos(\delta_2 -
\delta_1) = \cos \delta_1 \cos \delta_2 + \sin \delta_1 \sin \delta_2
$ in \eq{Re-expr} { and \eq{Re-exprb}} one realizes that these equations
are invariant, if the signs of $\cos \phi$, \dg, and $\cos
\delta_{1,2}$ are flipped simultaneously.  { Hence an overall
two-fold sign ambiguity persists and the sign of $\cos \phi$ remains
undetermined.}

The time evolution of the full three-angle distribution of the products
of the decay chain $B_s\to J/\psi[\to l^+l^-]\, \phi[\to K^+K^-]$ provides 
many interesting CP-violating observables \cite{ddf1,RF-ang}. 
{ The expressions for three-angle angular distributions can be
obtained by inserting (\ref{EQB}-\ref{EQEb}) into Eqs.~(64) and (70) of 
\cite{ddf1}.}

The situation is considerably simplified in the case of the one-angle
distribution, which takes the following form \cite{ddlr,ddf1}:
\begin{equation}\label{single}
\frac{d \Gamma (t)}{d \cos \Theta} \propto
(|A_0(t)|^2 + |A_{\|}(t)|^2)\,\frac{3}{8}\,(1 + \cos ^2 \Theta)
+ |A_{\perp}(t)|^2\,\frac{3}{4} \sin^2 \Theta \,.
\label{angel}
\end{equation}
Here $\Theta$ describes the angle between the decay direction of the $l^+$
and the $z$ axis in the $J/\psi$ rest frame; the $z$ axis is perpendicular
to the decay plane of $\phi\to K^+K^-$. With the help of this one-angle 
distribution, the observables $|A_0(t)|^2 + |A_{\|}(t)|^2$ and 
$|A_{\perp}(t)|^2$, as well as their CP conjugates, can be determined.
They provide the following CP asymmetries:
\begin{equation}\label{CP1}
\frac{ \left[|\overline{A}_0(t)|^2 + |\overline{A}_{\|}(t)|^2\right] -
       \left[|A_0(t)|^2 + |A_{\|}(t)|^2\right] 
       }{
       \left[|\overline{A}_0(t)|^2 +|\overline{A}_{\|}(t)|^2\right] 
     + \left[|A_0(t)|^2 + |A_{\|}(t)|^2\right] } 
\; = \; 
  \frac{ - \sin\phi \, \sin(\dm\,  t) }{ 
           \cosh (\dg\, t/2) - |\cos \phi | \sinh (|\dg|\, t/2) } 
\end{equation}
\begin{equation}\label{CP2}
\frac{ |\overline{A}_{\perp}(t)|^2 - |A_{\perp}(t)|^2 }{
       |\overline{A}_{\perp}(t)|^2 + |A_{\perp}(t)|^2 } 
\; = \;
\frac{ \sin\phi \, \sin( \dm\,  t) }{ 
           \cosh (\dg\, t/2) + |\cos \phi | \sinh (|\dg|\, t/2) } .
\end{equation}
In contrast to these CP-violating observables, untagged data samples are
sufficient to determine the following quantities:
\begin{eqnarray}
\lefteqn{\left[|A_0(t)|^2 + |A_{\|}(t)|^2\right]+\left[|\overline{A}_0(t)|^2 +
|\overline{A}_{\|}(t)|^2\right]}\nonumber\\
&&~~~~~=\; 2\, 
 \left[|A_0(0)|^2 + |A_{\|}(0)|^2\right] \, e^{-\Gamma t} \,
        \lt[ \cosh \frac{\dg\, t}{2} - |\cos \phi | 
        \sinh \frac{ |\dg|\, t}{2}
        \rt]\label{untag1}
\end{eqnarray}
\begin{equation}\label{untag2}
|A_{\perp}(t)|^2+|\overline{A}_{\perp}(t)|^2 
\; = \; 2\, |A_{\perp}(0)|^2 \, e^{-\Gamma t} \,
        \lt[ \cosh \frac{\dg\, t}{2} + |\cos \phi | 
        \sinh \frac{ |\dg|\, t}{2}
        \rt].
\end{equation}
Since $\phi$ is tiny in the Standard Model, a striking signal of
new-physics contributions to $B_s$--$\overline{B_s}$ mixing would be
provided by 
a sizeable $\sin \phi$ either from a fit of the tagged 
observables \eq{EQB} -- \eq{EQE}, \eq{EQBb} -- \eq{EQEb}, or from the
CP-violating asymmetries
{ in \eq{aph}}, (\ref{CP1}) and (\ref{CP2}), or if the untagged
observables (\ref{untag1}) and (\ref{untag2}) should depend on {\it
two} exponentials.  Note that in \eq{untag1} the coefficient of $\sinh
(|\dg|\, t/2)$ is always negative. Phrased differently, the
coefficient of the exponential $\exp(-(\Gamma+|\dg|/2) t) $ with the
larger rate is always larger than the coefficient of
$\exp(-(\Gamma-|\dg|/2) t) $.  In \eq{untag2} the situation is
reversed. This feature can be used as an experimental consistency
check, once $\dg \neq 0$ is established.

Let us finally note that the formalism developed in this subsection
applies also to the mode $B_s\to D_s^{\ast+}\,D_s^{\ast-}$, where the
subsequent decay of the $D_s^{\ast\pm}$-mesons is predominantly
electromagnetic, i.e.\ $D_s^{\ast\pm}\to D_s^{\pm}\gamma$. The
corresponding angular distribution can be found in \cite{ddlr,ddf1}. The
analysis of this decay requires the capability to detect photons and
appears to be considerably more challenging than that of $B_s\to
J/\psi\,\phi$, which is one of the ``gold-plated'' channels for
$B$-physics experiments at hadron machines. Higher $D_s$ resonances
exhibiting all-charged final states, for instance $D_{s1}(2536)^+\to
D^{\ast+}[\to D\pi^+]\, K$, may be more promising in this
respect~\cite{FD1}. If photon detection is not possible, one can still
distinguish $D_s^{\ast\pm}$'s from $D_s^\pm$'s through the energy
smearing associated with the escaped photon \cite{cp}. Then one can
use the lifetime method introduced in sect.~\ref{sec:both} to find the
CP-odd fraction $x$ ($\propto |A_{\perp}(0)|^2$) and the CP-even
fraction $1-x$ ($\propto |A_0(0)|^2 + |A_{\|}(0)|^2$) of the
$D_s^{\ast +} D_s^{\ast -}$ data sample through \eq{dgmx}. If
$x\neq1/2$ there are still non-vanishing CP asymmetries, although they
are diluted by $1- 2 x$. The corresponding formula for the CP asymmetry of this
weighted average of CP-even and CP-odd final states can readily be
obtained from \eq{EQB}--\eq{EQP} and \eq{EQBb}--\eq{EQPb}:
\begin{eqnarray}
\lefteqn{ 
\frac{  \Gamma ( \ov{B}_s (t) \to D_s^{\ast +} D_s^{\ast -} ) - 
        \Gamma ( B_s (t) \to D_s^{\ast +} D_s^{\ast -} ) 
       }{
        \Gamma ( \ov{B}_s (t) \to D_s^{\ast +} D_s^{\ast -} ) + 
        \Gamma ( B_s (t) \to D_s^{\ast +} D_s^{\ast -} ) 
        } \; = } \nn
&& \qquad \qquad \qquad \qquad \qquad \qquad \qquad 
  \frac{ - (1- 2x) \,  \sin\phi \, \sin(\dm \,  t)  }{ 
           \cosh (\dg\, t/2) \, - \, (1- 2 x)\, |\cos \phi | \sinh (|\dg|\,
        t/2) } . \label{dilu}
\end{eqnarray}
The same procedure can be done with the $D_s^{\pm} D_s^{\ast \mp}$
data sample or any other of the decay modes in Table~\ref{tab}. 

A complete angular analysis for the three-body decays in
Table~\ref{tab} is more involved than the analysis for $B_s \to \psi
\phi$.  For example in $B_s \to \psi { K_S K_S}$, the { $K_S$} pair does
not necessarily come from a vector resonance and could be in an S- or
D-wave or even have a larger angular momentum. In such cases one might
restrict oneself to a one-angle transversity analysis of \cite{dqstl}
or even satisfy oneself with the diluted asymmetries in \eq{dilu}.
\boldmath
%\section{The Unambiguous Determination of the\\ 
%$B_s$--$\overline{B_s}$ Mixing Phase}\label{sec:ambig}
\section{The Unambiguous Determination of $\phi$}\label{sec:ambig}
\unboldmath
%%Dunietz 
\newcommand{\sbarm}{
    \ensuremath{\stackrel{\scriptscriptstyle \! \lt( - \rt)}{S{}^{-}}}}
\newcommand{\sbarp}{
    \ensuremath{\stackrel{\scriptscriptstyle \! \lt( -\rt)}{S{}^{+}}}}
\newcommand{\pbm}{
    \ensuremath{\stackrel{\scriptscriptstyle  \lt( +\rt)}{-}}}
\newcommand{\mbp}{
    \ensuremath{\stackrel{\scriptscriptstyle  \lt( -\rt)}{+}}}
\newcommand{\gks}{\ensuremath{\Gamma_{short}}}
\newcommand{\gkl}{\ensuremath{\Gamma_{long}}}
\newcommand{\gk}{\ensuremath{\Gamma_K}}
\newcommand{\sinphi}{\ensuremath{\sin \phi}}
\newcommand{\cosphi}{\ensuremath{\cos \phi}}
\newcommand{\signcos}{\ensuremath{sign(\cos \phi)}}
\newcommand{\cosTwoBeta}{\ensuremath{\cos 2\widetilde{\beta}}}
\newcommand{\bsbsbar}{\ensuremath{
        \stackrel{\scriptscriptstyle \lt( -\rt)}{B}_s}}
\setlength{\unitlength}{1mm}
\newcommand{\subdecay}{
\parbox[b]{2truecm}{\vspace{-10mm}
\begin{picture}(18,13)
\put(0,1){\line(0,1){6}}
\put(0,1){\vector(1,0){18}}
\end{picture}
}}

While \sinphi\ can be measured by conventional methods, this section
shows that even \signcos\ can be determined.  That determination is
important { for various reasons.  It is not only necessary for a
complete extraction of magnitude and phase of the new physics
contributions to \bbms, $\phi$ must also be known to extract the CKM
angle $\gamma$ from $B_s \to D_s^{\pm} K^{\mp}$. Even if $\sin \phi$
is found to be consistent with zero, the determination of \signcos\ is
necessary to distinguish the Standard Model prediction
$\cosphi \simeq 1$} from  $\cosphi \simeq -1$.  In the advent of new physics,
\signcos\ completes our knowledge about $\phi$.  There are several
methods to extract $\cosphi$.

\paragraph{Method 1:} 
The previous section revealed that angular correlation studies of 
$B_s \to \psi \phi$ determine 
\begin{eqnarray}
\cos \delta_i \cosphi
\label{cosDeltaCosPhi} .
\end{eqnarray}
Once $sign( \cos \delta_i )$ is known, \signcos\ follows immediately.
The former can be deduced from theory, once first-principle
calculations of $\delta_i$ have progressed sufficiently~\cite{bbns}.
Alternatively, one can infer $sign( \cos \delta_i )$ from their
SU(3) counterparts occurring in $B_d \to \psi K^* [\to \pi^0 K_S],
\psi \rho^0, \psi \omega$ decays [denoted by $sign( \cos
\widehat{\delta}_i )$], as follows:

The angular correlations of those $B_d$ modes are sensitive to 
\cite{dqstl,ddf1}

$$\cos \widehat{\delta}_i \cos 2 \widetilde{\beta} . $$ 
By applying the SU(3) relation
%\footnote{
%Note that Ref.~\cite{ddfAmbig} implicitly assumes $\cosphi > 0$ when it
%determines $sign( \cosTwoBeta)$.  Once new physics effects are considered,
%\cosphi\ may be negative, and method 1 is still applicable.
%}
%end footnote
$$sign(\cos \delta_i) = sign(\cos \widehat{\delta}_i) ,$$
the relative
sign between \cosTwoBeta\ and \cosphi\ can be determined, but not yet
the absolute sign of \cosphi. That absolute sign can be determined,
since there are methods which extract the \bbmd\ phase $2
\widetilde{\beta}$ unambiguously, even in the presence of new physics
\cite{azimovRappaport,kayserLondon,kayser,soaresQuinn,charles}.  In
the absence of new physics, $\widetilde{\beta}$ equals the angle
$\beta$ of the CKM unitarity triangle.  In Ref.~\cite{ddfAmbig},
  basically the same approach was used to determine the sign of $\cos
  2\tilde\beta$. However, in that paper it was assumed that $\phi$ is
  negligibly small, as in the Standard Model. On the other hand, in
  method 1 we assume that $2\tilde\beta$ is known unambiguously,
  allowing the determination of $\cos\phi$. Using a theoretical input
  \cite{bbns} to determine $sign(\cos\delta_i)$ as noted above, the
  angular distribution of the $B_d\to J/\psi(\to
  l^+l^-)K^{*0}(\to\pi^0K_S)$ decay products considered in
  Ref.~\cite{ddfAmbig} also allows an unambiguous determination of
  $2\tilde\beta$ in the presence of $\phi\not=0$.

\paragraph{Method 2:} 
Consider certain three- (or $n$-) body modes $f$ that can be fed from both 
a $B_s$ and a $\ov B_s$, and where the \bsbsbar\ -decay amplitude is
a sum over a non-resonant contribution and several contributions via
quasi two-body modes. The strong phase variation can be modelled by
Breit-Wigners and is known, so that \cosphi\ can be extracted.  Such a method
was suggested in determining $\cos 2 \alpha$ 
and \cosTwoBeta\ in $B_d$ decays \cite{charles}.

An additional method can be found elsewhere~\cite{cascadeMethod}.

\section{Conclusions}\label{sec:concl}
In this paper we have addressed the experimental signatures of a
non-vanishing CP-violating phase $\phi$ in the \bbms\ amplitude. Since
$\phi$ is negligibly small in the Standard Model, but sizeable in many
of its extensions, it provides an excellent ground for the search of
new physics. We have discussed the determination of $\phi$ from both
untagged and tagged decays in CP-specific $B_s$ decay modes triggered
by the dominant quark level decays $\ov{b} \to \ov{c} c \ov{s}$ and
$\ov{b} \to \ov{c} u \ov{d}$. From lifetime measurements in these modes
one can find the product of $\cos \phi$ and the width difference $\dg$
in the $B_s$ system.  The previously proposed methods to separately
determine $|\dg|$ and $|\cos \phi|$ from untagged decay modes require
two-exponential fits to the time evolution of either flavour-specific
or CP-specific decay modes. In both cases terms of order $(\dg)^2$
must be experimentally resolved, which requires a substantially higher
statistics than needed to measure $\dg \cos \phi$. We have proposed a
new method to measure $|\dg|$ and $|\cos \phi|$, which only requires
lifetime fits to the collected data samples with double-charm final
states.  This method does not require sensitivity to ${\cal O}
((\dg)^2)$ terms.  It is based on the observation that the measurement
of $\dg$ from branching ratios discussed in \cite{ayopr} and performed
in \cite{aleph} is almost unaffected by new physics. These branching
ratios and $\dg \cos \phi$ obtained from the lifetime fits allow one to
solve for $|\dg|$ and $|\cos \phi|$.  In this context we have stressed
that the lifetime measurements also allow one to determine the size of the
CP-even and CP-odd components of $D_s^{\ast +} D_s^{\ast -}$ and
$D_s^{\pm} D_s^{\ast \mp}$ final states. This is relevant for
experiments which cannot detect photons { well enough} and therefore
cannot separate these components with angular analyses.  We have
further mentioned that a non-zero phase $\phi$ leads to tiny $\dm\, t$
oscillations in untagged data samples. This implies that { in
principle} the measurement of { CP violation in mixing from}
flavour-specific decays does not require tagging.

For the tagged analyses we have generalized the formulae for the CP
asymmetries to the case of a non-zero $\phi$. Here we have discussed in
detail the expressions needed for the angular analysis in $B_s \to
\psi \phi $ decays or other final states composed of two vector
particles. Finally we have shown how the discrete ambiguities in
$\phi$ encountered with the measurements of $|\cos \phi|$ and $\sin
\phi$ can be resolved and $\phi$ can be determined unambiguously.
This is important, even if $\sin \phi$ is found to be consistent
with zero, because it distinguishes the Standard Model case
$\phi\simeq 0$ from the case $\phi\simeq \pi$. If there are new
particles which couple to quarks with the same CKM elements as $W$
bosons, there can be new contributions to the \bbms\ amplitude
with larger magnitude, but opposite sign than the Standard Model box
diagram. In this case one encounters $\phi\simeq \pi$. This situation
can occur in multi-Higgs doublet models and in supersymmetric models
with flavour universality.  From a measurement of \dm\ alone the
contributions from the Standard Model and from new physics to the
\bbms\ amplitude cannot be separated.  The new contribution can only
be determined by combining the measurements of \dm\ and
$\phi$. Consider, for example, that \dm\ is measured in agreement with
the Standard Model prediction: the new physics contribution to \bbms\
then varies between 0 and twice the { Standard Model prediction},
if $\phi$ is varied between 0 and $\pm \pi$.
   
\section*{Acknowledgements}
I.D.\ and U.N.\ thank Farrukh Azfar, Stephen Bailey, Harry
Cheung, Petar Maksimovic, Matthew Martin, Hans-G\"unther Moser and
Christoph Paus for illuminating discussions.

\end{document}